\documentclass[aps,prd,reprint,longbibliography,%
	nofootinbib,eqsecnum,showpacs]{revtex4-1}
\usepackage{amssymb,amsmath,amsfonts}                     % American Mathematical Society packages for symbols, math, and fonts
\usepackage{mathtools}                                    % Extension to amsmath
\usepackage{bm}                                           % Bold symbols: \bm{}
\usepackage{graphicx}                                     % Extended Graphics
\graphicspath{{figures/}}
\usepackage{hyperref}                                     % Hypertext links [pagebackref] (must be loaded before cleveref)
\usepackage[capitalise,compress]{cleveref}                % Clever reference handling -- use with \cref{}
\usepackage{xstring}                                      % String manipulation
\usepackage{color}                                        % Color
\definecolor{DarkBlue}{rgb}{0.18,0.19,0.57}               % Define link color (color or xcolor package required)
\hypersetup{
	bookmarksnumbered=true,
	breaklinks=false, % default=false (do not break links between lines)
	citecolor=DarkBlue,
	colorlinks=true,
	linkcolor=DarkBlue,
	pdfauthor={Pantelis Pnigouras},
	pdfkeywords={},
	pdfstartpage=1, % default=1
	pdfsubject={},
	pdftitle={Gravitational-wave-driven tidal secular instability in neutron star binaries},
	urlcolor=DarkBlue,
}
\newcommand{\crefrangeconjunction}{--}                    % Put a dash instead of 'to' when linking equations in \cref{eq1,eq2,...}
\newcommand{\creflastconjunction}{, and\nobreakspace}     % Use of the serial comma for multiple references
\crefname{section}{Sec.}{Secs.}                           % section reference (mid-sentence)
\Crefname{section}{Section}{Sections}                     % section reference (sentence begining)
\crefformat{appsec}{the #2Appendix#3}                     % single appendix reference (mid-sentence) -- use with \label[appsec]{}
\Crefformat{appsec}{#2Appendix#3}                         % single appendix reference (sentence begining) --use with \label[appsec]{}
\newcommand{\refcite}[1]{%                                % Put Ref(s). before numeric-style citation(s) (mid-sentence; xstring package required)
	\begingroup
		\def\tempx{0}%
		\StrCount{#1}{,}[\tempx]%
		\ifnum\tempx > 0 
			Refs.~\cite{#1}%
		\else
			Ref.~\cite{#1}%
		\fi
	\endgroup
}
\newcommand{\Refcite}[1]{%                                % Put Reference(s) before numeric-style citation(s) (sentence beginning; xstring package required)
	\begingroup
		\def\tempx{0}%
		\StrCount{#1}{,}[\tempx]%
		\ifnum\tempx > 0 
			References~\cite{#1}%
		\else
			Reference~\cite{#1}%
		\fi
	\endgroup
}
\newcommand{\der}[2]{\frac{\mathrm{d}#1}{\mathrm{d}#2}}   % Ordinary derivative
\newcommand{\nder}[3]{\frac{\mathrm{d}^#1#2}{%            % Ordinary derivative of order n
	\mathrm{d}#3^#1}}
\newcommand{\parder}[2]{\frac{\partial#1}{\partial#2}}    % Partial derivative

\begin{document}
%
%-----------------------Title-----------------------------%
%
\title{Gravitational-wave-driven tidal secular instability in neutron star binaries}
\author{Pantelis Pnigouras}
	%\email{}
	\affiliation{Mathematical Sciences and STAG Research Centre, University of Southampton, Southampton SO17 1BJ, UK}
\date{\today}
%
%----------------------Abstract---------------------------%

\begin{abstract}
	We report the existence of a gravitational-wave-driven secular instability in neutron star binaries, acting on the equilibrium tide. The instability is similar to the classic Chandrasekhar-Friedman-Schutz (CFS) instability of normal modes and is active when the spin of the primary star exceeds the orbital frequency of the companion. Modeling the neutron star as a Newtonian $n=1$ polytrope, we calculate the instability time scale, which can be as low as a few seconds at small orbital separations but still larger than the inspiral time scale. The implications for orbital and spin evolution are also briefly explored, where it is found that the instability slows down the inspiral and decreases the stellar spin.
\end{abstract}

%--------------------PACS numbers-------------------------%

\pacs{}

\maketitle

%-----------------------------------------------------------------------------------------------------------------------------%
%%%%%%%%%%%%%%%%%%%%%%%%%%%%%%%%%%%%%%%%%%%%%%%%%%%%%%%%%%%%%%%%%%%%%%%%%%%%%%%%%%%%%%%%%%%%%%%%%%%%%%%%%%%%%%%%%%%%%%%%%%%%%%%
%-----------------------------------------------------------------------------------------------------------------------------%

\section{Introduction} \label{sec:Introduction}

Finite-size effects have been shown to play an important role during the late stages of a neutron star binary inspiral. Given that current gravitational-wave (GW) detectors rely mainly on searching for theoretically-predicted signals in their noisy data stream (matched filtering), the GW signal obtained by modeling the two stars as point particles may not be accurate enough due to phase errors induced by the tidal interaction (e.g., \refcite{Kochanek1992,BildstenCutler1992}). The significance of tidal effects on the GW signal and the binary evolution is determined by the tidal deformability of the stars, parametrized by the so-called tidal Love number \cite{Hinderer2008}, which depends on the neutron star equation of state, namely the equation of state of cold dense nuclear matter. Hence, the influence of the tidal interaction on the GW signal can be used to place constraints on the neutron star equation of state \cite{FlanaganHinderer2008}, something which was demonstrated already after the first detection of GWs from a neutron star binary \cite{AbbottEtAl2017m,AbbottEtAl2018d,RaithelEtAl2018,RadiceEtAl2018,DeEtAl2018}.

Another promising source of GWs and potential probe of the equation of state of supranuclear matter is neutron star instabilities. As discovered by Chandrasekhar \cite{Chandrasekhar1970} and rigorously proven by Friedman and Schutz \cite{FriedmanSchutz1978,FriedmanSchutz1978b}, certain oscillation modes in fast-rotating neutron stars are unstable to the emission of GWs. The instability occurs when a mode which is retrograde in the frame rotating with the star appears as prograde in the inertial frame. Then, the GWs emitted by the deformed star tend to increase the energy of the perturbation, causing the mode to grow on a secular time scale (for a review, see, e.g., \refcite{Andersson2003,Pnigouras2018}). This effect is more pronounced in large-scale perturbations, described by the star's fundamental modes ($f$-modes), which have no radial nodes and emit GWs more efficiently. In addition to polar modes, where density perturbations prevail and fast angular velocities are needed for the instability to develop, axial modes are also prone to the Chandrasekhar-Friedman-Schutz (CFS) instability. These are characterized mainly by perturbations of the fluid's horizontal velocity field and are caused by rotation itself ($r$-modes \cite{PapaloizouPringle1978}), which makes them CFS-unstable at all rotation rates (in the absence of viscosity) \cite{Andersson1998,FriedmanMorsink1998}.

Unstable oscillation modes are suitable for asteroseismology studies, in order to infer the neutron star equation of state, and can, in principle, generate large amounts of GWs \cite{LaiShapiro1995,OwenEtAl1998,DonevaEtAl2013,AlfordSchwenzer2014,DonevaEtAl2015}. Furthermore, the presence of the instability is expected to play a significant role in the evolution of nascent neutron stars and neutron stars in low-mass X-ray binaries (LMXBs), as shown by evolutionary studies \cite{Levin1999,AnderssonEtAl2000,PassamontiEtAl2013,BondarescuEtAl2007,BondarescuEtAl2009}, whereas it has been suggested as a possible explanation for the absence of pulsars spinning close to their break-up (mass-shedding) limit ($\sim 1\,\mathrm{kHz}$) \cite{Friedman1983,AnderssonEtAl1999,AnderssonEtAl1999b}. So far, Advanced LIGO and Virgo have not detected any evidence for such signals \cite{AbbottEtAl2017t,AbbottEtAl2019,AbbottEtAl2019b,AbbottEtAl2019c}.

In the present paper, we examine the stability of the tidal perturbation on a rotating star against the emission of GWs. To our knowledge, this problem has not been addressed by previous studies on gravitational radiation from tidally perturbed stars \cite{Mashhoon1973,Mashhoon1975,Mashhoon1977,Turner1977,Will1983} (however, see \refcite{HoLai1999}, where the resonant excitation of CFS-unstable modes from the tide is considered).

Assuming that the spin $\Omega$ of the primary star is aligned with the orbital angular velocity $\omega_\mathrm{orb}$ of its companion, the tidal perturbation induced on the primary will always be prograde in the inertial frame, but will appear retrograde in the rotating frame if $\Omega>\omega_\mathrm{orb}$. Using the equilibrium tide approximation, i.e., assuming that the perturbed star is always in hydrostatic equilibrium, we show that GWs generated by the tidally-deformed primary drive an instability, which develops on a secular time scale associated with the emission of GWs and has an impact on orbital and spin evolution.

We start by introducing the hydrodynamic equations describing a forced perturbation on the primary star, for which we derive an energy, as well as an energy rate which contains contributions from the varying tidal potential and from the gravitational radiation reaction force (\cref{subsec:Equation of motion,subsec:Perturbation energy}). Subsequently, we approximate the tidal perturbation with the equilibrium tide (\cref{subsec:The equilibrium tide}), which is computed analytically for a polytrope with index $n=1$ in \cref{sec:Equilibrium tide in an n=1 polytrope}. In \cref{sec:Equilibrium tide stability} we study the stability of the equilibrium tide against the emission of GWs, where we derive the instability criterion, calculate the instability growth time for a neutron star described by a polytropic equation of state with index $n=1$, and compare it to the inspiral time scale. In \cref{sec:Implications} we explore the implications of this instability for orbital and spin evolution, where we compute the corrections to the inspiral rate and the stellar spin for the same model. Finally, we summarize the main points and results and discuss some caveats and other considerations in \cref{sec:Summary and discussion}.

%-----------------------------------------------------------------------------------------------------------------------------%
%%%%%%%%%%%%%%%%%%%%%%%%%%%%%%%%%%%%%%%%%%%%%%%%%%%%%%%%%%%%%%%%%%%%%%%%%%%%%%%%%%%%%%%%%%%%%%%%%%%%%%%%%%%%%%%%%%%%%%%%%%%%%%%
%-----------------------------------------------------------------------------------------------------------------------------%

\section{The tidal perturbation} \label{sec:The tidal perturbation equations}

%-----------------------------------------------------------------------------------------------------------------------------%
%-----------------------------------------------------------------------------------------------------------------------------%

\subsection{Equation of motion} \label{subsec:Equation of motion}

We consider a star rotating with an angular velocity $\Omega$ (primary), perturbed by the tidal potential $U$ of a companion star. The primary is no longer in hydrostatic equilibrium, due to the tidal perturbation. The linearized (with respect to the perturbation) hydrodynamic equations for the primary, in the frame rotating with it, read
\begin{gather}
	\parder{\delta\rho}{t}+\nabla\cdot(\rho\delta\bm{v})=0, \label{perturbed continuity equation} \\
	\parder{\delta\bm{v}}{t}+2\bm{\Omega}\times\delta\bm{v}+\frac{\nabla\delta p}{\rho}-\frac{\nabla p}{\rho^2}\delta\rho+\nabla\delta\Phi+\nabla U = \bm{0}, \label{perturbed Euler equation} \\
	\nabla^2\delta\Phi=4\pi G\delta\rho, \label{perturbed Poisson equation} \\
	\nabla^2 U=0, \label{Laplace equation for tidal potential} \\
	\frac{\Delta p}{p}=\Gamma_1\frac{\Delta\rho}{\rho}, \label{perturbed equation of state}
\end{gather}
which are the (perturbed) continuity equation, Euler equation, Poisson equation, Laplace equation for the tidal potential, and equation of state, respectively. The symbols have their usual meanings: $\rho$ is the density, $p$ is the pressure, $\Phi$ is the gravitational potential, $\bm{v}$ is the velocity, whereas $t$ denotes time and $G$ is the gravitational constant. Eulerian and Lagrangian perturbations are denoted by $\delta$ and $\Delta$ respectively and are related by $\Delta f=\delta f+\left(\bm{\xi}\cdot\nabla\right)f$, where $\bm{\xi}$ is the displacement vector associated with the perturbation. The adiabatic exponent $\Gamma_1$ is defined as
\begin{equation}
	\Gamma_1=\left(\parder{\ln p}{\ln\rho}\right)_{x_p}, \label{adiabatic exponent}
\end{equation}
where $x_p$ denotes the proton fraction (i.e., the proton number density over the baryon number density), which generally varies throughout the star, but is considered as ``frozen'' during an orbital period $(\Delta x_p\approx 0)$, due to the slow time scales on which $\beta$ reactions operate \cite{ReiseneggerGoldreich1992}.\footnote{This assumption may not be valid at large orbital separations, but the current work is not concerned with this regime.}

Using the fact that $\delta\bm{v}=\dot{\bm{\xi}}$ (where the dot denotes the time derivative), \cref{perturbed continuity equation,perturbed Euler equation} are written as
\begin{equation}
	\delta\rho+\nabla\cdot(\rho\bm{\xi})=0 \label{perturbed continuity equation 2}
\end{equation}
and
\begin{equation}
	\ddot{\bm{\xi}}+2\bm{\Omega}\times\dot{\bm{\xi}}+\frac{\nabla\delta p}{\rho}-\frac{\nabla p}{\rho^2}\delta\rho+\nabla\delta\Phi+\nabla U = \bm{0} \label{perturbed Euler equation 2}
\end{equation}
respectively. Furthermore, using the relation between Lagrangian and Eulerian perturbations, \cref{perturbed equation of state} becomes
\begin{equation}
	\frac{\delta\rho}{\rho}=\frac{1}{\Gamma_1}\frac{\delta p}{p}-\bm{A}\cdot\bm{\xi}, \label{perturbed equation of state 2}
\end{equation}
where
\begin{equation}
	\bm{A}=\frac{\nabla\rho}{\rho}-\frac{1}{\Gamma_1}\frac{\nabla p}{p}. \label{Schwarzschild discriminant vector form}
\end{equation}
This is the Schwarzschild discriminant, with $|\bm{A}|\ne 0$ denoting the presence of buoyancy in the star (e.g., \refcite{UnnoEtAl1989}). In a star with no composition gradients $(x_p = \textrm{const.})$, $|\bm{A}|=0$ and perturbed fluid elements adjust instantaneously to the density of their surroundings.\footnote{Entropy gradients can also generate a nonzero buoyancy in a star, but are relevant mostly in newborn neutron stars, where the thermal pressure can be comparable to the degeneracy pressure \cite{KruegerEtAl2015}.} For later convenience, we rewrite the Euler equation \eqref{perturbed Euler equation 2} as
\begin{align}
	\ddot{\bm{\xi}} & +2\bm{\Omega}\times\dot{\bm{\xi}}+\nabla\left(\frac{p\Gamma_1}{\rho}\frac{\delta\rho}{\rho}+\delta\Phi+U\right) \notag \\
									& +\frac{p\Gamma_1}{\rho}\frac{\delta\rho}{\rho}\bm{A}+\frac{1}{\rho}\nabla(p\Gamma_1\bm{\xi}\cdot\bm{A}) = \bm{0}. \label{perturbed Euler equation 3}
\end{align}

%-----------------------------------------------------------------------------------------------------------------------------%
%-----------------------------------------------------------------------------------------------------------------------------%

\subsection{Perturbation energy} \label{subsec:Perturbation energy}

The rotating-frame energy associated with the perturbation is \cite{FriedmanSchutz1978b,SchenkEtAl2001}
\begin{equation}
	E=\frac{1}{2}\int\left[|\dot{\bm{\xi}}|^2+\bm{\xi}^*\cdot\bm{\mathcal{C}}(\bm{\xi})\right]\rho\mathrm{d}^3\bm{r}, \label{perturbation energy}
\end{equation}
where the operator $\bm{\mathcal{C}}$ is given by
\begin{equation}
	\bm{\mathcal{C}}(\bm{\xi})=\nabla\left(\frac{p\Gamma_1}{\rho}\frac{\delta\rho}{\rho}+\delta\Phi\right)+\frac{p\Gamma_1}{\rho}\frac{\delta\rho}{\rho}\bm{A}+\frac{1}{\rho}\nabla(p\Gamma_1\bm{\xi}\cdot\bm{A}). \label{operator C}
\end{equation}
Replacing the above in \cref{perturbation energy} and performing some integrations by parts, we get
\begin{align}
	E=\frac{1}{2}\int\Bigg\{ & \rho|\dot{\bm{\xi}}|^2+\frac{p\Gamma_1}{\rho}\frac{|\delta\rho|^2}{\rho}-\frac{1}{4\pi G}|\nabla\delta\Phi|^2 \notag \\
	                         &-p\Gamma_1\Bigg[(\bm{\xi}^*\cdot\bm{A})(\nabla\cdot\bm{\xi})+(\bm{\xi}\cdot\bm{A})(\nabla\cdot\bm{\xi}^*) \notag \\
	                         & \hspace{3em} +\frac{1}{\rho}(\bm{\xi}^*\cdot\bm{A})(\bm{\xi}\cdot\nabla\rho)\Bigg] \Bigg\}\mathrm{d}^3\bm{r}. \label{perturbation energy 2}
\end{align}
Due to the presence of the tidal potential, the perturbation energy changes at a rate (e.g., see \refcite{FriedmanSchutz1978b})
\begin{equation}
	\der{E}{t}=\mathrm{Re}\left[\int\dot{\bm{\xi}^*}\cdot\left(-\nabla U\right)\rho\mathrm{d}^3\bm{r}\right]. \label{perturbation energy rate}
\end{equation}

\Cref{perturbation energy rate} can be further supplemented with dissipation mechanisms, like GWs. In a Newtonian framework, GWs are implemented by introducing a potential which accounts for their emission by the perturbed star \citep{Thorne1969,IpserLindblom1991}. Then, the perturbation energy rate due to GW emission is
\begin{align}
	\left(\der{E}{t}\right)_\mathrm{GW} & =-\frac{1}{2}\sum_{l=2}^\infty\sum_{m=-l}^l(-1)^l N_l \notag \\
	\times\Bigg[ & \nder{{2l+1}}{}{t}\left(\delta D_l^m e^{-im\Omega t}\right)\der{\delta D_l^{*m}}{t}e^{im\Omega t} \notag \\
	+ & \nder{{2l+1}}{}{t}\left(\delta D_l^{*m} e^{im\Omega t}\right)\der{\delta D_l^m}{t}e^{-im\Omega t}\Bigg], \label{GW energy rate}
\end{align}
where
\begin{equation}
	N_l=\frac{4\pi G}{c^{2l+1}}\frac{(l+1)(l+2)}{l(l-1)\left[(2l+1)!!\right]^2}, \label{N_l}
\end{equation}
$c$ being the speed of light, and $\delta D_l^m$ are the mass multipole moments,\footnote{Typically, current multipole moments, accounting for gravitomagnetic effects, must also be included in \cref{GW energy rate} (see \refcite{Thorne1980}), but they are ignored here since we only have polar perturbations [see \cref{displacement vector}].} defined as
\begin{equation}
	\delta D_l^m=\int r^l \delta\rho\, Y_l^{*m}\mathrm{d}^3\bm{r}, \label{mass multipole moments}
\end{equation}
with $Y_l^m(\theta,\phi)$ denoting the spherical harmonic of degree $l$ and order $m$, defined in the rotating frame using a spherical coordinate system $(r,\theta,\phi)$.

The effects of GWs will be treated as secular, i.e., developing on a time scale much longer than the time scale associated with the perturbation. Then, \cref{GW energy rate} can be evaluated by using the solutions to the inviscid problem, namely the solutions to \cref{perturbed Euler equation 3} \citep{IpserLindblom1991}. This assumption will be shown to be valid in retrospect.

%-----------------------------------------------------------------------------------------------------------------------------%
%-----------------------------------------------------------------------------------------------------------------------------%

\subsection{The equilibrium tide} \label{subsec:The equilibrium tide}

The general solution for the tidal perturbation is often considered to comprise two parts: the \textit{equilibrium} and the \textit{dynamical} tides. The former corresponds to the instantaneous response of the primary to the tidal field of the companion, whereas the latter includes the resonant excitation of the primary's normal modes by the orbiting companion \cite{Zahn1966,Ogilvie2014}.

The equilibrium tide is simply obtained by assuming that the tidally perturbed star is in hydrostatic equilibrium, i.e., by setting the time derivatives in \cref{perturbed Euler equation 2} to zero. For simplicity, we will neglect the effects of rotation, in which case the eigenfunctions of the equilibrium tide are (e.g., \refcite{OgilvieLin2004})
\begin{gather}
	\delta p=-\rho(\delta\Phi+U), \label{pressure perturbation equilibrium tide} \\
	\delta\rho=\der{\rho}{r}\frac{\delta\Phi+U}{g}, \label{density perturbation equilibrium tide} \\
	\xi_r=-\frac{\delta\Phi+U}{g}, \label{displacement vector radial component equilibrium tide} \\
	\nabla\cdot\bm{\xi}=0, \label{displacement vector divergence equilibrium tide}
\end{gather}
and $\delta\Phi$ is given by
\begin{equation}
	\nabla^2\delta\Phi=4\pi G\der{\rho}{r}\frac{\delta\Phi+U}{g}, \label{perturbed Poisson equation equilibrium tide}
\end{equation}
where $g=\mathrm{d}\Phi/\mathrm{d}r$ and $\xi_r$ is the radial component of the displacement vector $\bm{\xi}$.

In order to express the tidal perturbation, we will use an inertial frame centered on the primary, with its $z$ axis parallel to the orbital angular momentum vector (generally not aligned with the primary's spin). Denoting the spherical coordinates of this frame as $(r,\theta',\phi')$, then the tidal potential is expanded in spherical harmonics as
\begin{equation}
	U=-\sum_{l=2}^\infty\sum_{m'=-l}^l\frac{GM'W_l^{m'} r^l}{D^{l+1}(t)}Y_l^{m'}(\theta',\phi')e^{-im'\Psi(t)}, \label{tidal potential}
\end{equation}
where $M'$ is the mass of the companion (treated as a point mass), $D(t)$ is the separation between the companion and the primary, $\Psi(t)$ is the orbital phase of the companion, and
\begin{align}
	W_l^{m'} & =\frac{4\pi}{2l+1}Y_l^{*m'}(\pi/2,0) \notag \\
	         & = (-)^{(l+m')/2}\left[\frac{4\pi}{2l+1}(l+m')!(l-m')!\right]^{1/2} \notag \\
	         & \hspace{2em} \times\left[2^l\left(\frac{l+m'}{2}\right)!\left(\frac{l-m'}{2}\right)!\right]^{-1}, \label{Wlm}
\end{align}
where $(-)^k=(-1)^k$, unless $k$ is not an integer, in which case it evaluates to zero \cite{PressTeukolsky1977}.

Considering a harmonic $(l,m')$ of the tidal potential and separating the radial, angular, and time dependence of the variables, then \cref{pressure perturbation equilibrium tide,density perturbation equilibrium tide,displacement vector radial component equilibrium tide} give the radial part of the corresponding eigenfunctions and \cref{perturbed Poisson equation equilibrium tide} becomes
\begin{equation}
	\frac{1}{r^2}\der{}{r}\left(r^2\der{\delta\Phi}{r}\right)-\frac{l(l+1)}{r^2}\delta\Phi = 4\pi G\der{\rho}{r}\frac{\delta\Phi+U}{g}=0 \label{perturbed Poisson equation equilibrium tide 2}
\end{equation}
(henceforth, the tidal potential $U$ and the perturbations will include only their radial dependence, i.e., their angular and time dependence shall be omitted). Replacing the displacement vector for polar perturbations, namely
\begin{equation}
	\bm{\xi}=\left[\xi_r,\xi_h\parder{}{{\theta'}},\frac{\xi_h}{\sin\theta'}\parder{}{{\phi'}}\right]Y_l^{m'}(\theta',\phi'), \label{displacement vector}
\end{equation}
in \cref{displacement vector divergence equilibrium tide}, we also obtain
\begin{equation}
	\xi_h=\frac{1}{l(l+1)r}\der{}{r}\left(r^2\xi_r\right). \label{displacement vector horizontal component equilibrium tide}
\end{equation}

For consistency, we will now express the tidal perturbation in the rotating frame used in \cref{subsec:Perturbation energy}. Let the rotating frame be described by the axes $(x,y,z)$, with the $z$-axis parallel to the primary's spin, and the inertial frame by $(x',y',z')$, with the $z'$-axis parallel to the orbital angular momentum. The two frames are related by the three Euler angles $(\alpha,\beta,\gamma)$, which are obtained as follows: rotate the inertial frame about the $z'$-axis by an angle $\alpha$ to obtain the frame $(x'_1,y'_1,z'_1=z')$; rotate the new frame about the $y'_1$-axis by an angle $\beta$ (spin--orbit inclination angle) to obtain the frame $(x'_2,y'_2=y'_1,z'_2)$ ---this is the rotating frame at $t=0$, so $z'_2=z$; finally, rotate the new frame about the $z$-axis by an angle $\gamma=\Omega t$, to obtain the rotating frame $(x,y,z)$.\footnote{This is the $z$-$y$-$z$ convention, using right-handed frames, with positive angles obtained by the right-handed screw rule \cite{SteinbornRuedenberg1973}.}

Then, the spherical harmonics of the inertial frame $Y_l^{m'}(\theta',\phi')$ are related to the spherical harmonics of the rotating frame $Y_l^m(\theta,\phi)$ as
\begin{equation}
	Y_l^{m'}(\theta',\phi')=\sum_{m=-l}^l D^{*(l)}_{m'm}(\alpha,\beta,\gamma)Y_l^m(\theta,\phi), \label{spherical harmonic transformation}
\end{equation}
where the (complex conjugate of the) Wigner $D$ function is given by
\begin{equation}
	D^{*(l)}_{m'm}=e^{im'\alpha}d^{(l)}_{m'm}(\beta)e^{im\gamma}, \label{Wigner D function}
\end{equation}
with
\begin{align}
	d^{(l)}_{m'm}(\beta) &= \left[(l+m)!(l-m)!(l+m')!(l-m')!\right]^{1/2} \notag \\
	&\hspace{-4em} \times \sum_k\frac{(-1)^{k+m'+m}\left(\cos\frac{\beta}{2}\right)^{2l+m-m'-2k}\left(\sin\frac{\beta}{2}\right)^{m'-m+2k}}{k!(l-m'-k)!(l+m-k)!(k+m'-m)!} \label{Wigner d function}
\end{align}
and the summation over $k$ runs over all integer values for which the factorial arguments are non-negative  \cite{SteinbornRuedenberg1973}.

%-----------------------------------------------------------------------------------------------------------------------------%
%%%%%%%%%%%%%%%%%%%%%%%%%%%%%%%%%%%%%%%%%%%%%%%%%%%%%%%%%%%%%%%%%%%%%%%%%%%%%%%%%%%%%%%%%%%%%%%%%%%%%%%%%%%%%%%%%%%%%%%%%%%%%%%
%-----------------------------------------------------------------------------------------------------------------------------%

\section{Equilibrium tide stability} \label{sec:Equilibrium tide stability}

According to the classic CFS instability for normal modes in rotating stars, an oscillation on the star becomes unstable to the emission of gravitational radiation when its inertial-frame frequency changes sign \cite{FriedmanSchutz1978,FriedmanSchutz1978b}. For a mode with a harmonic dependence $e^{i(m\phi+\omega t)}$, where $\omega$ is its rotating-frame frequency, \cref{GW energy rate} takes the familiar form \cite{IpserLindblom1991}
\begin{equation}
	\left(\der{E}{t}\right)_\mathrm{GW}=-\omega(\omega-m\Omega)\sum_{l=l_\mathrm{min}}^\infty N_l (\omega-m\Omega)^{2l}\left|\delta D_l^m\right|^2, \label{normal mode GW energy rate}
\end{equation}
where $l_\mathrm{min}=\max(2,|m|)$. \Cref{normal mode GW energy rate} shows that $\dot{E}_\mathrm{GW}>0$ if and only if $\omega(\omega-m\Omega)<0$. Thus, the onset of the instability occurs when the angular velocity of the star is such that the inertial-frame frequency of the mode, $\omega_\mathrm{in}\equiv\omega-m\Omega$, becomes zero. The instability can only affect retrograde modes $(m>0)$, i.e., modes propagating against the rotation of the star, which, under the influence of rotation, appear as prograde in the inertial frame.

In the following, we will study the conditions under which the CFS instability affects the equilibrium tide. For simplicity, we will assume that the spin of the primary is aligned with the orbital angular momentum $(\beta=0)$. Then, since $d^{(l)}_{m'm}(0)=\delta_{m'm}$ (where $\delta_{m'm}$ is Kronecker's delta), \cref{spherical harmonic transformation} gives the anticipated result\footnote{The constant phase difference $\alpha$ between the two frames can be set to zero.}
\begin{equation}
	Y_l^m(\theta,\phi')=Y_l^m(\theta,\phi)e^{im\Omega t}. \label{spherical harmonic transformation for beta=0}
\end{equation}
Hence, the harmonic time dependence of the equilibrium tide in the rotating frame is $e^{im[\Omega t-\Psi(t)]}$.

%-----------------------------------------------------------------------------------------------------------------------------%
%-----------------------------------------------------------------------------------------------------------------------------%

\subsection{Circular orbit} \label{subsec:Circular orbit}

In the simple case of a static circular orbit, we have $\Psi(t)=\omega_\mathrm{orb} t$, where $\omega_\mathrm{orb}$ is the orbital angular velocity, and the binary separation $D=\textrm{\textit{const}}$. Then, the time dependence of the equilibrium tide in the rotating frame becomes $e^{im(\Omega-\omega_\mathrm{orb})t}$ and \cref{GW energy rate}, for a specific harmonic $(l,m)$ of the tide, gives
\begin{multline}
	\left(\der{E}{t}\right)_\mathrm{GW}=-N_l \, m^2 \omega_\mathrm{orb}(\omega_\mathrm{orb}-\Omega) (m\omega_\mathrm{orb})^{2l} \\
	\times \left[\int_0^R \der{\rho}{r}\frac{\delta\Phi+U}{g} r^{l+2}\mathrm{d}r\right]^2. \label{equilibrium tide GW energy rate}
\end{multline}
This shows that $\dot{E}_\mathrm{GW}>0$ if
\begin{equation}
	\Omega>\omega_\mathrm{orb} \label{instability criterion circular orbit}
\end{equation}
or, replacing $\omega_\mathrm{orb}$ from Kepler's law and normalizing $\Omega$ to the Kepler (mass-shedding) limit for spherical stars, $\Omega_\mathrm{K}=\sqrt{GM/R^3}$ (where $R$ is the primary's radius)\footnote{A more accurate approximation of the Kepler limit in Newtonian stars can be obtained by using the Roche model and is given by $\Omega_\mathrm{K}=(2/3)^{3/2} \sqrt{GM/R^3}$ \cite{ShapiroTeukolsky1983}.} \cite{ShapiroTeukolsky1983}, the instability criterion takes the elegant form
\begin{equation}
	F(\Omega,D,M')\equiv \left(\frac{\Omega}{\Omega_\mathrm{K}}\right)\left(\frac{D}{R}\right)^{3/2}-\left(1+\frac{M'}{M}\right)^{1/2}>0. \label{instability criterion circular orbit 2}
\end{equation}

The time scale $\tau_\mathrm{GW}$ associated with damping (or growth) due to GWs is \cite{IpserLindblom1991}
\begin{equation}
	\tau_\mathrm{GW}=\frac{2E}{\dot{E}_\mathrm{GW}}, \label{GW time scale}
\end{equation}
where the energy of the $(l,m)$ harmonic of the equilibrium tide can be obtained by replacing the eigenfunctions of \cref{subsec:The equilibrium tide} in \cref{perturbation energy 2}, which gives
\begin{multline}
	E = \frac{1}{2}[m(\Omega-\omega_\mathrm{orb})]^2\int_0^R\Bigg\{\left(\frac{\delta\Phi+U}{g}\right)^2 + \frac{1}{l(l+1)} \\
	\times\left[-\frac{r}{g}\left(\der{\delta\Phi}{r}+\frac{lU}{r}\right)+\frac{\delta\Phi+U}{g}\left(\der{\ln g}{\ln r}-2\right)\right]^2\Bigg\}\rho r^2\mathrm{d}r \\
	-\frac{1}{8\pi G}\int_0^\infty\left[\left(r\der{\delta\Phi}{r}\right)^2+l(l+1)(\delta\Phi)^2\right]\mathrm{d}r \\
	+\frac{1}{2}\int_0^R\left(\frac{\delta\Phi+U}{g}\right)^2\der{\ln\rho}{r}\der{p}{r}r^2\mathrm{d}r. \label{equilibrium tide energy}
\end{multline} 

\begin{figure}[t]
	\includegraphics[width=\columnwidth]{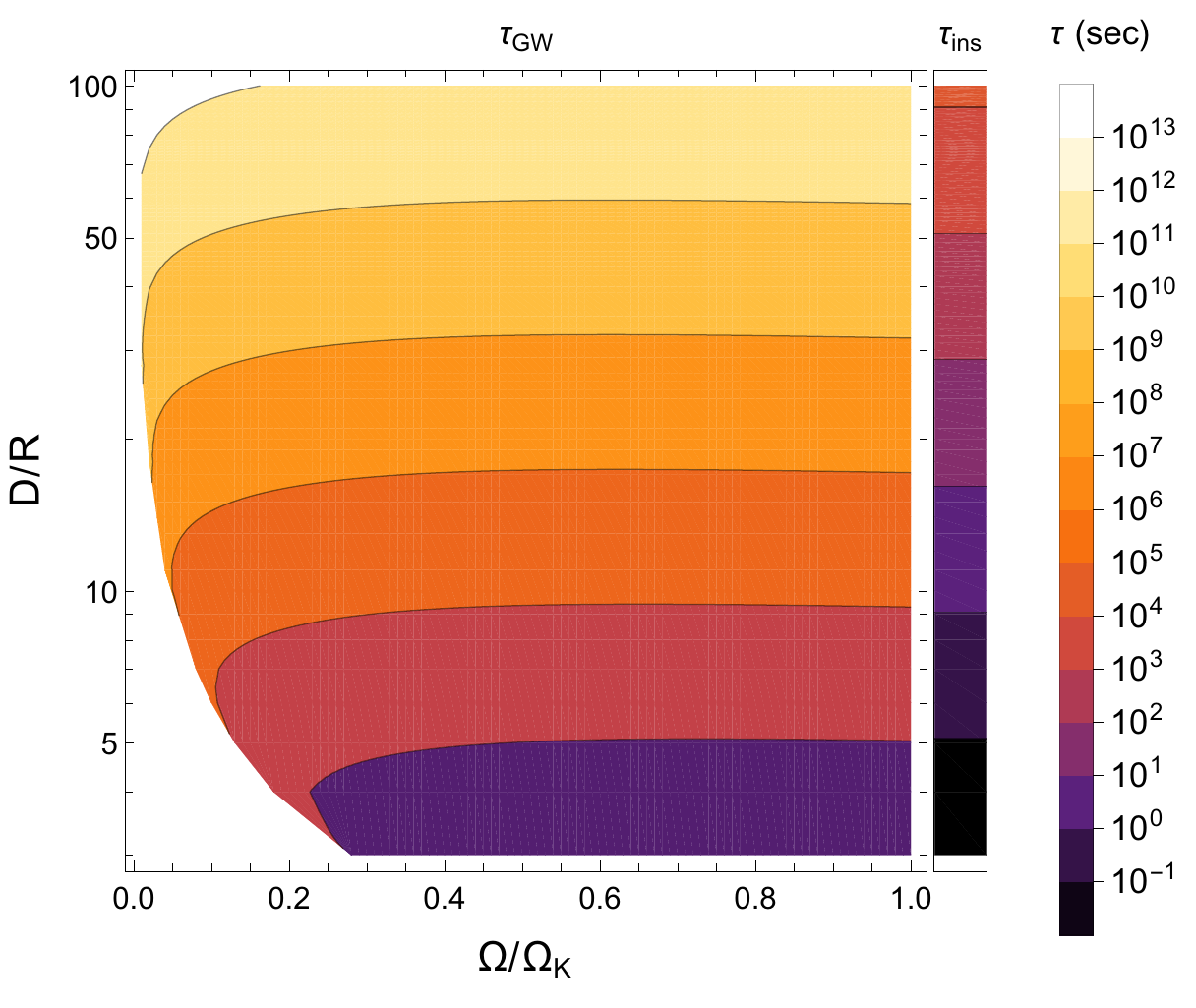}
	\caption{Instability growth time $\tau_\mathrm{GW}$ (in sec) as a function of the angular velocity $\Omega$ of the primary (normalized to the Kepler limit $\Omega_\mathrm{K}$; $x$-axis) and the orbital distance $D$ (normalized to the primary's radius $R$; $y$-axis). The primary is a neutron star described by an $n=1$ polytropic equation of state, with $M=1.4\, M_\odot$ and $R=10\,\mathrm{km}$, orbited by an equal mass companion $(M'=M)$. The inspiral time scale $\tau_\mathrm{ins}$ is also shown, as a function of the orbital distance, for comparison.}
	\label{fig:instability_and_inspiral_timescale_contour_l2m2_delta1}
\end{figure}

Assuming that the star is described by a polytrope with index $n=1$, we can now evaluate the instability time scale $\tau_\mathrm{GW}$ for the $l=2$ components of the equilibrium tide, the eigenfunctions of which can be obtained analytically (as shown in \cref{sec:Equilibrium tide in an n=1 polytrope}). For a neutron star with $M=1.4\,M_\odot$ (where $M_\odot$ is the solar mass) and $R=10\,\mathrm{km}$ the instability growth time is given by
\begin{multline}
	\tau_\mathrm{GW}=\Bigg[8.5\times 10^{-4}\left(\frac{D}{R}\right)^9 F^{-1}(\Omega,D,M') \\
											  +2.3\times 10^{-3}\left(\frac{D}{R}\right)^6 F(\Omega,D,M')\Bigg]\left(1+\frac{M'}{M}\right)^{-5/2}. \label{GW time scale for n=1 polytrope}
\end{multline}
In this model, $\nu_\mathrm{K}=\Omega_\mathrm{K}/2\pi\approx 2.2\,\mathrm{kHz}$. \Cref{GW time scale for n=1 polytrope} is plotted in \cref{fig:instability_and_inspiral_timescale_contour_l2m2_delta1} for $F(\Omega,D,M')>0$ (i.e., where the instability is active) and $M'=M$. For reasonable values of the mass ratio $M'/M$, the time scale is not significantly affected.

%-----------------------------------------------------------------------------------------------------------------------------%
%-----------------------------------------------------------------------------------------------------------------------------%

\subsection{Inspiral} \label{subsec:Inspiral}

The orbital motion of two stars generates GWs, gradually shrinking the binary's orbit and eventually leading to its coalesence. For two point-like stars in a quasi-circular orbit the orbital separation changes due to quadrupole emission of GWs as (e.g., \refcite{Maggiore2008})
\begin{equation}
	\der{D}{t}=-\frac{64G^3}{5c^5}\frac{M M'(M+M')}{D^3}, \label{orbital separation evolution}
\end{equation}
whereas the orbital phase $\Psi$ evolves according to
\begin{equation}
	\der{\Psi}{t}=\omega_\mathrm{orb}=\sqrt{\frac{G(M+M')}{D^3}}, \label{orbital frequency evolution}
\end{equation}
from which we also get
\begin{equation}
	\der{\omega_\mathrm{orb}}{t}=-\frac{3}{2}\frac{\dot{D}}{D}\omega_\mathrm{orb}. \label{orbital frequency derivative}
\end{equation}
The orbit is quasi-circular in the sense that the inspiral rate $|\dot{D}|/D$ is much smaller than $\omega_\mathrm{orb}$.

The energy of the tidal perturbation changes due to the orbital shrinking according to \cref{perturbation energy rate} which, evaluated for a certain harmonic of the equilibrium tide, gives
\begin{equation}
	\left(\der{E}{t}\right)_\mathrm{ins}=-(l+1)\frac{\dot{D}}{D}\frac{GM' W_l^m}{D^{l+1}}\int_0^R \der{\rho}{r}\frac{\delta\Phi+U}{g} r^{l+2}\mathrm{d}r \label{equilibrium tide perturbation energy rate}
\end{equation}
(note that the tidal potential $U$ and the perturbations are now functions not only of $r$, but also of $D(t)$, which will be henceforth implied). Using the relation between the mass multipole moments and the tidal Love number $k_l$ (e.g., \refcite{YipLeung2017}), \cref{equilibrium tide perturbation energy rate} can also be written as
\begin{equation}
	\left(\der{E}{t}\right)_\mathrm{ins}=-2k_l R^{2l+1} \frac{(2l+1)(l+1)}{4\pi G}\frac{\dot{D}}{D}\left(\frac{GM' W_l^m}{D^{l+1}}\right)^2. \label{equilibrium tide perturbation energy rate 2}
\end{equation}

In a similar manner, the energy rate of the tidal perturbation due to GW emission is obtained from \cref{GW energy rate} as
\begin{multline}
	\left(\der{E}{t}\right)_\mathrm{GW}= \;(-1)^{l+1} N_l \left[\int_0^R \der{\rho}{r}\frac{\delta\Phi+U}{g} r^{l+2}\mathrm{d}r\right]^2 \\
	\times \mathrm{Re}\left[(f^*-im\Omega)\left(f+\der{}{t}\right)^{2l}f\right], \label{equilibrium tide GW energy rate 2}
\end{multline}
where $f(t)=-(l+1)\dot{D}/D-im\omega_\mathrm{orb}$. For $l\sim m\neq 0$ we have $|\mathrm{Re}(f)|\ll|\mathrm{Im}(f)|$. From \cref{orbital frequency derivative} we also see that $\dot{\omega}_\mathrm{orb}\ll\omega_\mathrm{orb}^2$, which implies that derivatives of $\mathrm{Im}(f)$ can be neglected. Then, we recover \cref{equilibrium tide GW energy rate,instability criterion circular orbit}, albeit with a time dependence on $\omega_\mathrm{orb}$ and the perturbation variables. This can also be expressed in terms of the Love number as
\begin{multline}
	\left(\der{E}{t}\right)_\mathrm{GW}= -N_l(m\omega_\mathrm{orb})^{2l+2}\left(1-\frac{\Omega}{\omega_\mathrm{orb}}\right) \\
	\times \left(2k_l R^{2l+1}\frac{2l+1}{4\pi G}\frac{GM' W_l^m}{D^{l+1}}\right)^2. \label{equilibrium tide GW energy rate 3}
\end{multline}

The significance of the instability can be assessed by comparing the growth time $\tau_\mathrm{GW}$ to the inspiral time scale, given by
\begin{equation}
	\tau_\mathrm{ins}=\frac{D}{|\dot{D}|}=\frac{5c^5}{64G^3}\frac{D^4}{MM'(M+M')} \label{inspiral time scale}
\end{equation}
which, for $M=1.4\, M_\odot$ and $R=10\,\mathrm{km}$, evaluates as
\begin{equation}
	\tau_\mathrm{ins} = 2.95\times 10^{-4} \left(\frac{D}{R}\right)^4 \left[\frac{M'}{M}\left(1+\frac{M'}{M}\right)\right]^{-1}. \label{inspiral time scale 2}
\end{equation}
This is also plotted in \cref{fig:instability_and_inspiral_timescale_contour_l2m2_delta1}, alongside the instability growth time. Both from \cref{fig:instability_and_inspiral_timescale_contour_l2m2_delta1} and from a direct comparison between \cref{GW time scale for n=1 polytrope,inspiral time scale 2} it becomes evident that the inspiral time scale is shorter than the time required for the instability to develop for all values of $D$ and $\Omega$.

%-----------------------------------------------------------------------------------------------------------------------------%
%%%%%%%%%%%%%%%%%%%%%%%%%%%%%%%%%%%%%%%%%%%%%%%%%%%%%%%%%%%%%%%%%%%%%%%%%%%%%%%%%%%%%%%%%%%%%%%%%%%%%%%%%%%%%%%%%%%%%%%%%%%%%%%
%-----------------------------------------------------------------------------------------------------------------------------%

\section{Implications} \label{sec:Implications}

Using some simple arguments, we will present the implications of this instability of the equilibrium tide on the orbital and spin evolution, restricting ourselves to the quadrupole components $(l=2)$. The energy rate of the tidal perturbation, denoted below as $\dot{E}_\mathrm{tide}$, is given by \cref{equilibrium tide perturbation energy rate 2,equilibrium tide GW energy rate 3} and can be written as
\begin{equation}
	\dot{E}_\mathrm{tide}=\epsilon\dot{E}_\mathrm{tide}^{(1)}+\epsilon^2\dot{E}_\mathrm{tide}^{(2)}, \label{equilibrium tide quadrupole components total energy rate}
\end{equation}
where
\begin{equation}
	\epsilon=2k_2\left(1+\frac{M'}{M}\right)\left(\frac{R}{D}\right)^5 \label{epsilon}
\end{equation}
and $\dot{E}_\mathrm{tide}^{(1)},\,\dot{E}_\mathrm{tide}^{(2)}$, which will be given below, contain the contributions of the inspiral and of GW emission to the energy rate of the tidal perturbation. 

For a binary system in a quasi-circular orbit, where the tidal deformation of the primary (but not of the companion) is taken into account, the GW power emitted from the system is \cite{Chau1976,Clark1977}
\begin{equation}
	\dot{E}_\mathrm{GW}=\dot{E}_\mathrm{GW}^{(0)}+\epsilon\dot{E}_\mathrm{GW}^{(1)}+\epsilon^2 \dot{E}_\mathrm{GW}^{(2)}, \label{equilibrium tide quadrupole components GW power}
\end{equation}
where
\begin{align}
	\dot{E}_\mathrm{GW}^{(1)} & =2\dot{E}_\mathrm{GW}^{(0)}, \label{equilibrium tide quadrupole components GW power first order} \\
	\dot{E}_\mathrm{GW}^{(2)} & =\dot{E}_\mathrm{GW}^{(0)}, \label{equilibrium tide quadrupole components GW power second order}
\end{align}
and
\begin{equation}
	\dot{E}_\mathrm{GW}^{(0)}=-\frac{32 G^4 (MM')^2(M+M')}{5c^5 D^5}, \label{equilibrium tide quadrupole components GW power zeroth order}
\end{equation}
which is the point-mass limit \cite{PetersMathews1963}. Note that $\dot{E}_\mathrm{GW}$ is not to be confused with the contribution of GW emission to the energy rate of the tidal perturbation [\cref{GW energy rate,equilibrium tide GW energy rate,equilibrium tide GW energy rate 2,equilibrium tide GW energy rate 3}], which is here contained in $\dot{E}_\mathrm{tide}$ (see below).

Expanding the orbital energy losses in a similar way, we have
\begin{equation}
	\dot{E}_\mathrm{orb}=\dot{E}_\mathrm{orb}^{(0)}+\epsilon \dot{E}_\mathrm{orb}^{(1)} + \epsilon^2\dot{E}_\mathrm{orb}^{(2)}. \label{orbital energy rate}
\end{equation}
At zeroth order in $\epsilon$, we simply get
\begin{equation}
	\dot{E}_\mathrm{orb}^{(0)} = \dot{E}_\mathrm{GW}^{(0)}, \label{orbital energy rate zeroth order}
\end{equation}
from which we obtain the orbital decay in the point-mass limit, given by \cref{orbital separation evolution}. Then, using this in \cref{equilibrium tide perturbation energy rate 2}, we get
\begin{equation}
	\dot{E}_\mathrm{tide}^{(1)} =-\frac{6M'}{M+M'}\dot{E}_\mathrm{GW}^{(0)}. \label{equilibrium tide quadrupole components total energy rate first order}
\end{equation}
Hence, the first-order correction to the orbital energy rate is
\begin{equation}
	\dot{E}_\mathrm{orb}^{(1)} = \dot{E}_\mathrm{GW}^{(1)}-\dot{E}_\mathrm{tide}^{(1)} \label{orbital energy rate first order}
\end{equation}
or, replacing \cref{equilibrium tide quadrupole components GW power first order,equilibrium tide quadrupole components total energy rate first order},
\begin{equation}
	\dot{E}_\mathrm{orb}^{(1)}=\left(2+\frac{6M'}{M+M'}\right)\dot{E}_\mathrm{GW}^{(0)}. \label{orbital energy rate first order 2}
\end{equation}
This can be used to calculate the first-order correction to the inspiral rate due to the tidal deformation of the primary. If $\dot{D}=\dot{D}^{(0)}+\epsilon \dot{D}^{(1)}$ (where $\dot{D}^{(0)}$ is the point-mass result), we find that
\begin{equation}
	\dot{D}^{(1)}=\left(2+\frac{6M'}{M+M'}\right)\dot{D}^{(0)}, \label{orbital separation evolution first order}
\end{equation}
namely, the inspiral is accelerated, as expected \cite{Kochanek1992}.

At second order, both the inspiral and GW emission contribute to the tidal perturbation energy rate. The contribution of GW emission, obtained from \cref{equilibrium tide GW energy rate 3}, is
\begin{equation}
	\dot{E}^{(2)}_\mathrm{tide,\,GW}=\left(1-\frac{\Omega}{\omega_\mathrm{orb}}\right)\dot{E}^{(0)}_\mathrm{GW}. \label{equilibrium tide quadrupole components total energy rate second order GWs}
\end{equation}
The contribution of the inspiral is found by replacing the first-order correction to the inspiral rate [\cref{orbital separation evolution first order}] back to \cref{equilibrium tide perturbation energy rate 2}, which gives
\begin{equation}
	\dot{E}^{(2)}_\mathrm{tide,\,ins}=-\frac{6M'}{M+M'}\left(2+\frac{6M'}{M+M'}\right)\dot{E}^{(0)}_\mathrm{GW}. \label{equilibrium tide quadrupole components total energy rate second order inspiral}
\end{equation}
Thus, the second-order correction to the orbital energy rate is
\begin{equation}
	\dot{E}_\mathrm{orb}^{(2)} = \dot{E}_\mathrm{GW}^{(2)}-\dot{E}_\mathrm{tide}^{(2)}-\dot{E}^{(2)}_\mathrm{bg}, \label{orbital energy rate second order}
\end{equation}
where we also added possible changes in the energy of the background (unperturbed) star, $\dot{E}^{(2)}_\mathrm{bg}$. Replacing \cref{equilibrium tide quadrupole components GW power second order,equilibrium tide quadrupole components total energy rate second order GWs,equilibrium tide quadrupole components total energy rate second order inspiral}, we get
\begin{equation}
	\dot{E}_\mathrm{orb}^{(2)}+\dot{E}^{(2)}_\mathrm{bg}=\left[\frac{\Omega}{\omega_\mathrm{orb}}+\frac{6M'}{M+M'}\left(2+\frac{6M'}{M+M'}\right)\right]\dot{E}^{(0)}_\mathrm{GW}. \label{orbital and stellar energy rate second order}
\end{equation}

In order to proceed with \cref{orbital and stellar energy rate second order}, we need to also consider the emission of angular momentum from the binary system. For GW emission from the tidal bulge, for which the second order term in \cref{equilibrium tide quadrupole components GW power} is responsible, angular momentum is emitted at a rate $\epsilon^2\dot{J}^{(2)}_\mathrm{GW}$, given by \cite{Thorne1980,Chugunov2019}
\begin{equation}
	\dot{J}^{(2)}_\mathrm{GW}=\frac{\dot{E}^{(2)}_\mathrm{GW}}{\omega_\mathrm{orb}}. \label{equilibrium tide quadrupole components GW angular momentum rate second order}
\end{equation}
Likewise, for the orbit we have
\begin{equation}
	\dot{J}_\mathrm{orb}=\frac{\dot{E}_\mathrm{orb}}{\omega_\mathrm{orb}}. \label{orbital angular momentum rate}
\end{equation}
The angular momentum rate associated with the tidal perturbation can be obtained by computing the torque applied on the star by the tidal force, as well as by the gravitational radiation reaction force, namely the force that accounts for GWs (see \refcite{IpserLindblom1991}). The total torque (along the $z$-axis) is given by \cite{Lai1994}
\begin{equation}
	T=\int\delta\rho\, \bm{e}_z\cdot\left(\bm{r}\times\bm{F}^*\right)\mathrm{d}^3\bm{r}, \label{torque}
\end{equation}
where $\bm{F}$ is the corresponding force (e.g., the tidal force is $\bm{F}=-\nabla U$). Evaluation of \cref{torque} shows that $\dot{J}^{(2)}_\mathrm{tide,\,ins}=0$, which is expected, since we are only considering the equilibrium tide where the tidal bulge is always aligned with the companion.\footnote{Thus, in this case, there is no dynamical tidal lag due to the inspiral (see \refcite{Lai1994}). Also, since we ignore viscosity, there is no viscosity-induced tidal lag either \cite{Kochanek1992,BildstenCutler1992,Lai1994}.} In addition, we get
\begin{equation}
	\dot{J}^{(2)}_\mathrm{tide,\,GW}=\frac{\dot{E}^{(2)}_\mathrm{tide,\,GW}}{\omega_\mathrm{orb}-\Omega}\;. \label{equilibrium tide quadrupole components angular momentum rate second order}
\end{equation}

\begin{figure*}
	\begin{tabular}{cc}
		\includegraphics[width=\columnwidth]{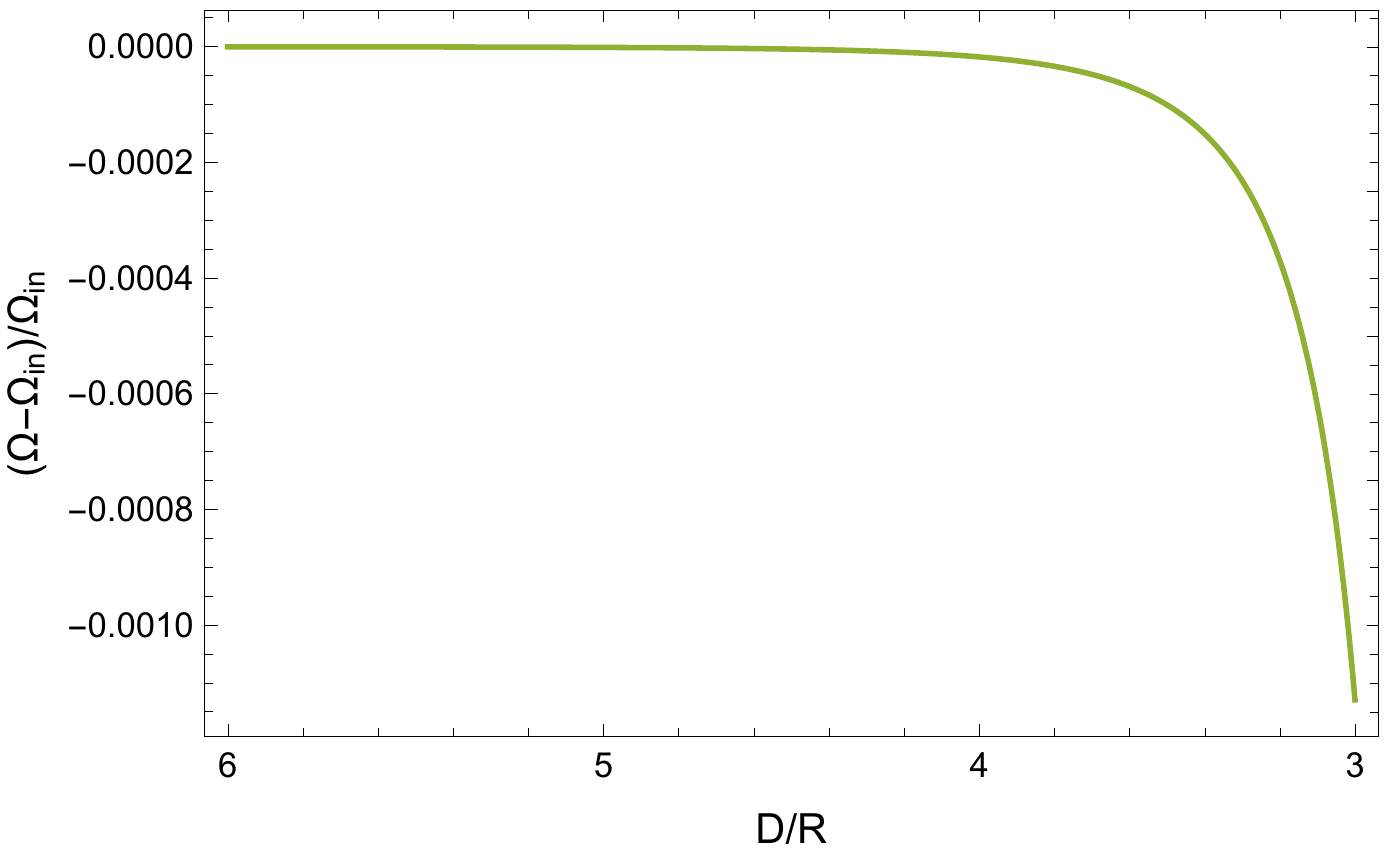} & \includegraphics[width=\columnwidth]{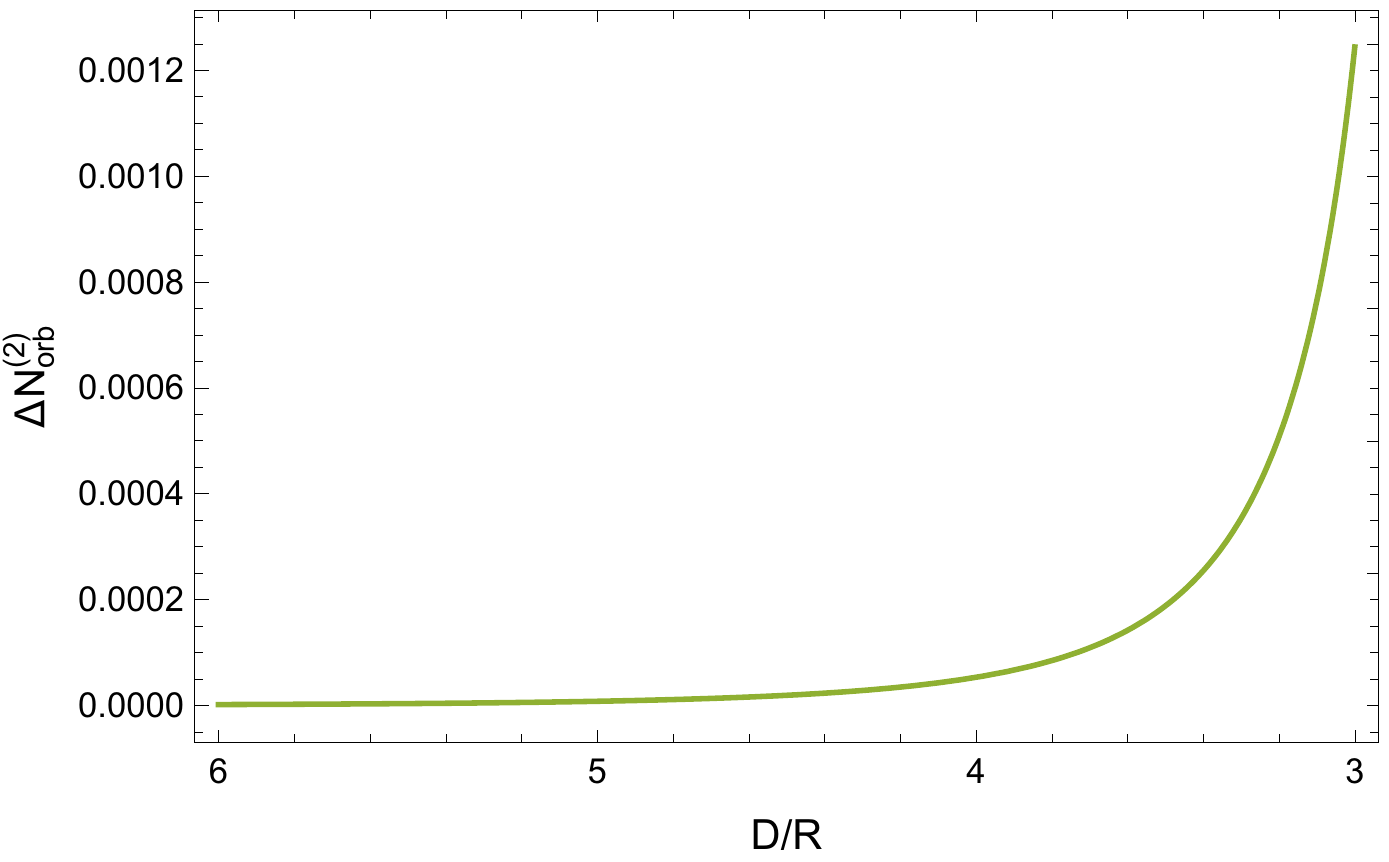}
	\end{tabular}
	\caption{Change in the primary's spin $\Omega$, relatively to its initial value $\Omega_\mathrm{in}$ (\textit{left}), and second-order contribution to the number of orbital cycles $\Delta N_\mathrm{orb}^{(2)}$ (\textit{right}), plotted against the orbital separation $D$ (normalized to the primary's radius $R$). The primary is a neutron star described by an $n=1$ polytropic equation of state, with $M=1.4\, M_\odot$ and $R=10\,\mathrm{km}$, orbited by an equal mass companion $(M'=M)$. The evolution is started at an orbital separation $D_\mathrm{in}=100\,R$ and the angular velocity of the primary is set to $\Omega_\mathrm{in}=0.3\,\Omega_\mathrm{K}$, where $\Omega_\mathrm{K}$ is the Kepler limit (for spherical stars). The binary merger is taken to occur at $3 R$.}
	\label{fig:Omega_and_DeltaNSecondOrder_versus_D_delta1_OmegaInitial30pc}
\end{figure*}

Hence, for the angular momentum emission from the system we have
\begin{equation}
	\frac{\dot{E}^{(2)}_\mathrm{GW}}{\omega_\mathrm{orb}}=\frac{\dot{E}^{(2)}_\mathrm{orb}}{\omega_\mathrm{orb}}+\frac{\dot{E}^{(2)}_\mathrm{tide,\,GW}}{\omega_\mathrm{orb}-\Omega}+\frac{\dot{E}^{(2)}_\mathrm{bg}}{\Omega}, \label{angular momentum conservation}
\end{equation}
where we also replaced $\dot{E}^{(2)}_\mathrm{bg}=\Omega\dot{J}^{(2)}_\mathrm{bg}$. Using \cref{equilibrium tide quadrupole components GW power second order,equilibrium tide quadrupole components total energy rate second order GWs}, we finally obtain
\begin{equation}
	\frac{\dot{E}^{(2)}_\mathrm{orb}}{\omega_\mathrm{orb}}=-\frac{\dot{E}^{(2)}_\mathrm{bg}}{\Omega}, \label{orbital and stellar energy rate second order 2}
\end{equation}
which, replaced in \cref{orbital and stellar energy rate second order}, gives
\begin{equation}
	\dot{E}^{(2)}_\mathrm{orb}=\frac{\displaystyle\frac{\Omega}{\omega_\mathrm{orb}}+\frac{6M'}{M+M'}\left(2+\frac{6M'}{M+M'}\right)}{\displaystyle 1-\frac{\Omega}{\omega_\mathrm{orb}}}\dot{E}^{(0)}_\mathrm{GW} \label{orbital energy rate second order 2}
\end{equation}
and
\begin{equation}
	\dot{E}^{(2)}_\mathrm{bg}=\frac{\displaystyle\frac{\Omega}{\omega_\mathrm{orb}}+\frac{6M'}{M+M'}\left(2+\frac{6M'}{M+M'}\right)}{\displaystyle 1-\frac{\omega_\mathrm{orb}}{\Omega}}\dot{E}^{(0)}_\mathrm{GW}. \label{stellar energy rate second order}
\end{equation}
From \cref{orbital energy rate second order 2,stellar energy rate second order} we may now obtain the second-order correction to the inspiral rate $\epsilon^2\dot{D}^{(2)}$ and the background star's spin derivative, respectively, as
\begin{equation}
	\dot{D}^{(2)}=-\frac{\displaystyle F(\Omega,D,M')+\left(1+\frac{M'}{M}\right)^{1/2}+\mathcal{T}(M')}{F(\Omega,D,M')}\dot{D}^{(0)} \label{orbital separation evolution second order}
\end{equation}
and
\begin{multline}
	\frac{\dot{\Omega}}{\Omega_\mathrm{K}}=-\frac{\displaystyle F(\Omega,D,M')+\left(1+\frac{M'}{M}\right)^{1/2}+\mathcal{T}(M')}{F^2(\Omega,D,M')} \\
	\times\frac{2\tilde{E}_\mathrm{tide}(\Omega,D,M')}{\tau_\mathrm{GW}(\Omega,D,M')}\left(\frac{D}{R}\right)^{3/2}, \label{spin evolution}
\end{multline}
where
\begin{equation}
	\mathcal{T}(M')=12\frac{M'}{M}\left(1+4\frac{M'}{M}\right)\left(1+\frac{M'}{M}\right)^{-3/2} \label{T}
\end{equation}
and
\begin{equation}
	\tilde{E}_\mathrm{tide}=\frac{E_\mathrm{tide}}{I\Omega_\mathrm{K}^2}, \label{equilibrium tide quadrupole components energy normalised}
\end{equation}
with $E_\mathrm{tide}$ being the energy of the quadrupole components of the equilibrium tide [see \cref{equilibrium tide energy}] and $I$ being the background star's moment of inertia. For a polytrope with $n=1$, $M=1.4\,M_\odot$ and $R=10\,\mathrm{km}$, we have
\begin{multline}
	\tilde{E}_\mathrm{tide} = \, 2.03\left(\frac{M'}{M}\right)^2\left(\frac{D}{R}\right)^{-9}F^2(\Omega,D,M') \\
	+0.75\left(\frac{M'}{M}\right)^2\left(\frac{D}{R}\right)^{-6}, \label{equilibrium tide quadrupole components energy normalised for n=1 polytrope}
\end{multline}
whereas $\tau_\mathrm{GW}$ is given by \cref{GW time scale for n=1 polytrope}.

\Cref{orbital separation evolution second order} predicts that, when the instability is active [$F(\Omega,D,M')>0$], the inspiral is decelerated, which has also been shown to occur during the resonant excitation of CFS-unstable normal modes by the tide \cite{HoLai1999}. On the other hand, according to \cref{spin evolution}, the spin of the (unperturbed) primary is decreasing when the instability occurs ($\tau_\mathrm{GW}>0$), in accordance with \refcite{Chugunov2019}, where the spin evolution equation is derived for unstable $r$-modes.

Solving the equations for $\dot{\Omega}$ and $\dot{D}=\dot{D}^{(0)}+\epsilon\dot{D}^{(1)}+\epsilon^2\dot{D}^{(2)}$, for the same model used above, we find that the change in the spin of the primary is negligible, as shown in \cref{fig:Omega_and_DeltaNSecondOrder_versus_D_delta1_OmegaInitial30pc}. In the same figure, we also plot the second-order contribution to the number of orbital cycles, $\Delta N_\mathrm{orb}^{(2)}$, defined as \cite{Lai1994}
\begin{equation}
	N_\mathrm{orb}=\int\frac{\omega_\mathrm{orb}}{2\pi}\frac{\mathrm{d}D}{\dot{D}}=\int\frac{\omega_\mathrm{orb}}{2\pi}\frac{\mathrm{d}D}{\dot{D}^{(0)}+\epsilon\dot{D}^{(1)}}+\Delta N_\mathrm{orb}^{(2)}, \label{orbital cycles}
\end{equation}
with $N_\mathrm{orb}$ being the total number of orbits at second order. We see that the correction is positive, as expected from the discussion above, and accumulates very close to merger, as is the correction for the spin. However, this too is unimportant, at least for the parameters and the model considered here.

%-----------------------------------------------------------------------------------------------------------------------------%
%%%%%%%%%%%%%%%%%%%%%%%%%%%%%%%%%%%%%%%%%%%%%%%%%%%%%%%%%%%%%%%%%%%%%%%%%%%%%%%%%%%%%%%%%%%%%%%%%%%%%%%%%%%%%%%%%%%%%%%%%%%%%%%
%-----------------------------------------------------------------------------------------------------------------------------%

\vspace{-1.2em}
\section{Summary and discussion} \label{sec:Summary and discussion}
\vspace{-.7em}

We have shown that the equilibrium tide, namely the instantaneous hydrostatic response of a star to the tidal field of its companion, is unstable to the emission of gravitational radiation if the spin of the star exceeds the orbital angular velocity of the companion [\cref{instability criterion circular orbit,instability criterion circular orbit 2}]. When this condition is fulfilled, the tidal perturbation, which is always prograde in the inertial frame, becomes retrograde in the frame rotating with the star. Then, the emission of GWs from the tidal bulge tends to increase the energy of the tidal perturbation on a secular time scale. This mechanism shares the same principles with the classic Chandrasekhar-Friedman-Schutz (CFS) instability for normal modes in rotating stars.

The instability growth time was calculated for a neutron star with $M=1.4\,M_\odot$ and $R=10\,\mathrm{km}$, described by a polytropic equation of state with a polytropic index $n=1$ [\cref{GW time scale for n=1 polytrope}]. For this model, the eigenfunctions of the equilibrium tide can be derived analytically. As seen in \cref{fig:instability_and_inspiral_timescale_contour_l2m2_delta1}, the instability is active in a very large part of the parameter space and the growth time varies by many orders of magnitude throughout the inspiral, depending mainly on the orbital separation and less on the spin of the star. Even though it can become as low as a few seconds very close to coalesence, the instability growth is always slower than the inspiral---at least for the chosen model.

Finally, the implications of the instability for orbital and spin evolution are explored, making use of some basic energy arguments. The corrections to the inspiral rate due to the tidal deformation and the emission of GWs from the tide are found [\cref{orbital separation evolution first order,orbital separation evolution second order}], along with the change in the stellar spin [\cref{spin evolution}]. It is demonstrated that, when the instability is active, the emission of GWs from the tide slows down the inspiral, which has also been shown to be a consequence of the resonant excitation of CFS-unstable normal modes by the tide \cite{HoLai1999}. Meanwhile, the angular velocity of the star decreases, evolving in a similar fashion as when under the influence of the classic CFS instability of normal modes \cite{Chugunov2019}. In \cref{fig:Omega_and_DeltaNSecondOrder_versus_D_delta1_OmegaInitial30pc} it is shown that, for the same model used above, the effects of the instability on orbital and spin evolution start becoming relevant---but still negligible---only very close to merger.

It should be noted that the rotation rates required for the instability to develop are unlikely to occur during the last stages of the binary evolution. However, near coalesence, the inspiral rate becomes too fast (plunge phase) for the quasi-cicrular orbit approximation to be accurate. Moreover, at this stage, the equilibrium tide approximation may also not be valid. If the tidal frequency is larger than the frequency of convective motions in the star (Brunt-V\"ais\"al\"a frequency\footnote{The Brunt-V\"ais\"al\"a frequency $N$ is given by $N^2=-gA$, where $g$ is the local gravitational acceleration and $A$ is the Schwarzschild discriminant [\cref{Schwarzschild discriminant vector form}].}), then the star responds effectively like a barotrope, in which case the equilibrium tide approximation fails \cite{TerquemEtAl1998,GoodmanDickson1998,Ogilvie2014}.

Furthermore, an important element which, for simplicity, has been neglected here is viscosity. Even though it makes the orbital evolution more dynamical, by introducing a lag between the orbital motion of the companion and the primary's response, it has been shown not to significantly affect the orbital and spin evolution \cite{Kochanek1992,BildstenCutler1992}. Nevertheless, it is expected to supress the instability and shrink the parameter space where it is active, as in the case of CFS-unstable modes \cite{IpserLindblom1991}.

From the above, it seems that the instability is purely of conceptual interest. Even so, there are some cases which might be worth considering in the future, like neutron stars described by stiffer equations of state, corresponding to larger deformabilities \cite{Hinderer2008}, or systems in which the neutron star has a much more massive companion.

%--------------------Acknowledgments----------------------%

\vspace{-1.3em}
\begin{acknowledgments}
	\vspace{-.7em}
	Support from STFC via grant ST/R00045X/1 is gratefully acknowledged. The author would like to thank N.~Andersson and D.~I.~Jones, for enlightening discussions and helpful suggestions, as well as the referee, for useful comments.
\end{acknowledgments}

%-----------------------Appendix--------------------------%

\renewcommand{\appendixname}{APPENDIX}                    % Capitalise appendix name

\appendix*

% Use the following with appendix* to avoid hyperref
% 	package warning (Original: \def\thesection{\unskip},
% 	contained in appendix definition)
\def\thesection{\texorpdfstring{\unskip}{}}

\let\oldsection\section                                   % Capitalise appendix section title
\def\section#1{\oldsection{%
	\texorpdfstring{\MakeUppercase{#1}}{Appendix: #1}%
}}

\section{Equilibrium tide in an \texorpdfstring{$\bm{n=1}$}{n=1} polytrope} \label[appsec]{sec:Equilibrium tide in an n=1 polytrope}

The primary is described by a polytropic equation of state with polytropic index $n=1$, namely $p=K\rho^2$, where $K$ is the polytropic constant. For a star with mass $M$ and radius $R$, the Lane-Emden equations (e.g., see \refcite{ShapiroTeukolsky1983}) give
\begin{gather}
	K=\frac{2GR^2}{\pi}, \label{polytropic constant for n=1 polytrope} \\
	\rho_c=\frac{\pi M}{4R^3}, \label{central density for n=1 polytrope} \\
	\rho=\rho_c\frac{\sin(\alpha r)}{\alpha r}, \label{density for n=1 polytrope} \\
	g=\frac{4\pi G\rho_c}{\alpha^3 r^2}\left[\sin(\alpha r)-\alpha r\cos(\alpha r)\right], \label{gravitational acceleration for n=1 polytrope}
\end{gather}
where $\alpha=\pi/R$.

%K=4.25537\times 10^{4}\,\mathrm{g}^{-1}\;\mathrm{cm}^5\,\mathrm{s}^{-2}
%\rho_c=2.11498\times 10^{15}\;\mathrm{g}\,\mathrm{cm}^{-3}

Replacing in \cref{perturbed Poisson equation equilibrium tide 2} and combining with \cref{Laplace equation for tidal potential} we get
\begin{equation}
	\der{}{r}\left[r^2\der{(\delta\Phi+U)}{r}\right]+\left[(\alpha r)^2-l(l+1)\right](\delta\Phi+U)=0 \label{perturbed Poisson equation equilibrium tide for n=1 polytrope}
\end{equation}
(the tidal potential $U$ and the perturbations include only their radial dependence), the solutions to which are the spherical Bessel functions $j_l(\alpha r)$ and $y_l(\alpha r)$ (e.g., \refcite{AbramowitzStegun1972}). Retaining the solution that is regular at the origin and matching at the surface with the external solution for $\delta\Phi$, we obtain
\begin{equation}
	\delta\Phi=\left\{
	\begin{array}{lr}
		\displaystyle\left[\frac{2l+1}{\pi j_{l-1}(\pi)}j_l(\alpha r)-\left(\frac{r}{R}\right)^l\right]U(R), & r\leq R, \\[2em]
		\displaystyle\frac{j_{l+1}(\pi)}{j_{l-1}(\pi)}\left(\frac{R}{r}\right)^{l+1}U(R), & r>R,
	\end{array}
	\right. \label{gravitational potential perturbation equilibrium tide for n=1 polytrope}
\end{equation}
with $j_l$ given by
\begin{equation}
	j_l(z)=z^l\left(-\frac{1}{z}\der{}{z}\right)^l\frac{\sin z}{z}. \label{spherical Bessel functions of the first kind}
\end{equation}
The equations above, together with \cref{tidal potential} for the tidal potential, can now be used to obtain the eigenfunctions of the equilibrium tide, as presented in \cref{subsec:The equilibrium tide}.

%--------------------Bibliography-------------------------%
%
\bibliography{articles,books}

\begin{thebibliography}{63}%
\makeatletter
\providecommand \@ifxundefined [1]{%
 \@ifx{#1\undefined}
}%
\providecommand \@ifnum [1]{%
 \ifnum #1\expandafter \@firstoftwo
 \else \expandafter \@secondoftwo
 \fi
}%
\providecommand \@ifx [1]{%
 \ifx #1\expandafter \@firstoftwo
 \else \expandafter \@secondoftwo
 \fi
}%
\providecommand \natexlab [1]{#1}%
\providecommand \enquote  [1]{``#1''}%
\providecommand \bibnamefont  [1]{#1}%
\providecommand \bibfnamefont [1]{#1}%
\providecommand \citenamefont [1]{#1}%
\providecommand \href@noop [0]{\@secondoftwo}%
\providecommand \href [0]{\begingroup \@sanitize@url \@href}%
\providecommand \@href[1]{\@@startlink{#1}\@@href}%
\providecommand \@@href[1]{\endgroup#1\@@endlink}%
\providecommand \@sanitize@url [0]{\catcode `\\12\catcode `\$12\catcode
  `\&12\catcode `\#12\catcode `\^12\catcode `\_12\catcode `\%12\relax}%
\providecommand \@@startlink[1]{}%
\providecommand \@@endlink[0]{}%
\providecommand \url  [0]{\begingroup\@sanitize@url \@url }%
\providecommand \@url [1]{\endgroup\@href {#1}{\urlprefix }}%
\providecommand \urlprefix  [0]{URL }%
\providecommand \Eprint [0]{\href }%
\providecommand \doibase [0]{http://dx.doi.org/}%
\providecommand \selectlanguage [0]{\@gobble}%
\providecommand \bibinfo  [0]{\@secondoftwo}%
\providecommand \bibfield  [0]{\@secondoftwo}%
\providecommand \translation [1]{[#1]}%
\providecommand \BibitemOpen [0]{}%
\providecommand \bibitemStop [0]{}%
\providecommand \bibitemNoStop [0]{.\EOS\space}%
\providecommand \EOS [0]{\spacefactor3000\relax}%
\providecommand \BibitemShut  [1]{\csname bibitem#1\endcsname}%
\let\auto@bib@innerbib\@empty
%</preamble>
\bibitem [{\citenamefont {Kochanek}(1992)}]{Kochanek1992}%
  \BibitemOpen
  \bibfield  {author} {\bibinfo {author} {\bibfnamefont {C.~S.}\ \bibnamefont
  {Kochanek}},\ }\bibfield  {title} {\enquote {\bibinfo {title} {Coalescing
  binary neutron stars},}\ }\href {\doibase 10.1086/171851} {\bibfield
  {journal} {\bibinfo  {journal} {Astrophys. J.}\ }\textbf {\bibinfo {volume}
  {398}},\ \bibinfo {pages} {234--247} (\bibinfo {year} {1992})}\BibitemShut
  {NoStop}%
\bibitem [{\citenamefont {Bildsten}\ and\ \citenamefont
  {Cutler}(1992)}]{BildstenCutler1992}%
  \BibitemOpen
  \bibfield  {author} {\bibinfo {author} {\bibfnamefont {L.}~\bibnamefont
  {Bildsten}}\ and\ \bibinfo {author} {\bibfnamefont {C.}~\bibnamefont
  {Cutler}},\ }\bibfield  {title} {\enquote {\bibinfo {title} {Tidal
  interactions of inspiraling compact binaries},}\ }\href {\doibase
  10.1086/171983} {\bibfield  {journal} {\bibinfo  {journal} {Astrophys. J.}\
  }\textbf {\bibinfo {volume} {400}},\ \bibinfo {pages} {175--180} (\bibinfo
  {year} {1992})}\BibitemShut {NoStop}%
\bibitem [{\citenamefont {Hinderer}(2008)}]{Hinderer2008}%
  \BibitemOpen
  \bibfield  {author} {\bibinfo {author} {\bibfnamefont {T.}~\bibnamefont
  {Hinderer}},\ }\bibfield  {title} {\enquote {\bibinfo {title} {Tidal Love
  Numbers of Neutron Stars},}\ }\href {\doibase 10.1086/533487} {\bibfield
  {journal} {\bibinfo  {journal} {Astrophys. J.}\ }\textbf {\bibinfo {volume}
  {677}},\ \bibinfo {pages} {1216--1220} (\bibinfo {year} {2008})},\ \Eprint
  {http://arxiv.org/abs/0711.2420} {arXiv:0711.2420} \BibitemShut {NoStop}%
\bibitem [{\citenamefont {Flanagan}\ and\ \citenamefont
  {Hinderer}(2008)}]{FlanaganHinderer2008}%
  \BibitemOpen
  \bibfield  {author} {\bibinfo {author} {\bibfnamefont {{\'E}.~{\'E}.}\
  \bibnamefont {Flanagan}}\ and\ \bibinfo {author} {\bibfnamefont
  {T.}~\bibnamefont {Hinderer}},\ }\bibfield  {title} {\enquote {\bibinfo
  {title} {Constraining neutron-star tidal Love numbers with gravitational-wave
  detectors},}\ }\href {\doibase 10.1103/PhysRevD.77.021502} {\bibfield
  {journal} {\bibinfo  {journal} {Phys. Rev. D}\ }\textbf {\bibinfo {volume}
  {77}},\ \bibinfo {eid} {021502} (\bibinfo {year} {2008})},\ \Eprint
  {http://arxiv.org/abs/0709.1915} {arXiv:0709.1915} \BibitemShut {NoStop}%
\bibitem [{\citenamefont {Abbott}\ \emph
  {et~al.}(2017{\natexlab{a}})\citenamefont {Abbott} \emph
  {et~al.}}]{AbbottEtAl2017m}%
  \BibitemOpen
  \bibfield  {author} {\bibinfo {author} {\bibfnamefont {B.~P.}\ \bibnamefont
  {Abbott}} \emph {et~al.},\ }\bibfield  {title} {\enquote {\bibinfo {title}
  {GW170817: Observation of Gravitational Waves from a Binary Neutron Star
  Inspiral},}\ }\href {\doibase 10.1103/PhysRevLett.119.161101} {\bibfield
  {journal} {\bibinfo  {journal} {Phys. Rev. Lett.}\ }\textbf {\bibinfo
  {volume} {119}},\ \bibinfo {eid} {161101} (\bibinfo {year}
  {2017}{\natexlab{a}})},\ \Eprint {http://arxiv.org/abs/1710.05832}
  {1710.05832} \BibitemShut {NoStop}%
\bibitem [{\citenamefont {Abbott}\ \emph {et~al.}(2018)\citenamefont {Abbott}
  \emph {et~al.}}]{AbbottEtAl2018d}%
  \BibitemOpen
  \bibfield  {author} {\bibinfo {author} {\bibfnamefont {B.~P.}\ \bibnamefont
  {Abbott}} \emph {et~al.},\ }\bibfield  {title} {\enquote {\bibinfo {title}
  {GW170817: Measurements of Neutron Star Radii and Equation of State},}\
  }\href {\doibase 10.1103/PhysRevLett.121.161101} {\bibfield  {journal}
  {\bibinfo  {journal} {Phys. Rev. Lett.}\ }\textbf {\bibinfo {volume} {121}},\
  \bibinfo {eid} {161101} (\bibinfo {year} {2018})},\ \Eprint
  {http://arxiv.org/abs/1805.11581} {arXiv:1805.11581 [gr-qc]} \BibitemShut
  {NoStop}%
\bibitem [{\citenamefont {Raithel}\ \emph {et~al.}(2018)\citenamefont
  {Raithel}, \citenamefont {{\"O}zel},\ and\ \citenamefont
  {Psaltis}}]{RaithelEtAl2018}%
  \BibitemOpen
  \bibfield  {author} {\bibinfo {author} {\bibfnamefont {C.~A.}\ \bibnamefont
  {Raithel}}, \bibinfo {author} {\bibfnamefont {F.}~\bibnamefont {{\"O}zel}}, \
  and\ \bibinfo {author} {\bibfnamefont {D.}~\bibnamefont {Psaltis}},\
  }\bibfield  {title} {\enquote {\bibinfo {title} {Tidal Deformability from
  GW170817 as a Direct Probe of the Neutron Star Radius},}\ }\href {\doibase
  10.3847/2041-8213/aabcbf} {\bibfield  {journal} {\bibinfo  {journal}
  {Astrophys. J.}\ }\textbf {\bibinfo {volume} {857}},\ \bibinfo {eid} {L23}
  (\bibinfo {year} {2018})},\ \Eprint {http://arxiv.org/abs/1803.07687}
  {arXiv:1803.07687 [astro-ph.HE]} \BibitemShut {NoStop}%
\bibitem [{\citenamefont {Radice}\ \emph {et~al.}(2018)\citenamefont {Radice},
  \citenamefont {Perego}, \citenamefont {Zappa},\ and\ \citenamefont
  {Bernuzzi}}]{RadiceEtAl2018}%
  \BibitemOpen
  \bibfield  {author} {\bibinfo {author} {\bibfnamefont {D.}~\bibnamefont
  {Radice}}, \bibinfo {author} {\bibfnamefont {A.}~\bibnamefont {Perego}},
  \bibinfo {author} {\bibfnamefont {F.}~\bibnamefont {Zappa}}, \ and\ \bibinfo
  {author} {\bibfnamefont {S.}~\bibnamefont {Bernuzzi}},\ }\bibfield  {title}
  {\enquote {\bibinfo {title} {GW170817: Joint Constraint on the Neutron Star
  Equation of State from Multimessenger Observations},}\ }\href {\doibase
  10.3847/2041-8213/aaa402} {\bibfield  {journal} {\bibinfo  {journal}
  {Astrophys. J.}\ }\textbf {\bibinfo {volume} {852}},\ \bibinfo {eid} {L29}
  (\bibinfo {year} {2018})},\ \Eprint {http://arxiv.org/abs/1711.03647}
  {arXiv:1711.03647 [astro-ph.HE]} \BibitemShut {NoStop}%
\bibitem [{\citenamefont {De}\ \emph {et~al.}(2018)\citenamefont {De},
  \citenamefont {Finstad}, \citenamefont {Lattimer}, \citenamefont {Brown},
  \citenamefont {Berger},\ and\ \citenamefont {Biwer}}]{DeEtAl2018}%
  \BibitemOpen
  \bibfield  {author} {\bibinfo {author} {\bibfnamefont {S.}~\bibnamefont
  {De}}, \bibinfo {author} {\bibfnamefont {D.}~\bibnamefont {Finstad}},
  \bibinfo {author} {\bibfnamefont {J.~M.}\ \bibnamefont {Lattimer}}, \bibinfo
  {author} {\bibfnamefont {D.~A.}\ \bibnamefont {Brown}}, \bibinfo {author}
  {\bibfnamefont {E.}~\bibnamefont {Berger}}, \ and\ \bibinfo {author}
  {\bibfnamefont {C.~M.}\ \bibnamefont {Biwer}},\ }\bibfield  {title} {\enquote
  {\bibinfo {title} {Tidal Deformabilities and Radii of Neutron Stars from the
  Observation of GW170817},}\ }\href {\doibase 10.1103/PhysRevLett.121.091102}
  {\bibfield  {journal} {\bibinfo  {journal} {Phys. Rev. Lett.}\ }\textbf
  {\bibinfo {volume} {121}},\ \bibinfo {eid} {091102} (\bibinfo {year}
  {2018})},\ \Eprint {http://arxiv.org/abs/1804.08583} {arXiv:1804.08583
  [astro-ph.HE]} \BibitemShut {NoStop}%
\bibitem [{\citenamefont {{Chandrasekhar}}(1970)}]{Chandrasekhar1970}%
  \BibitemOpen
  \bibfield  {author} {\bibinfo {author} {\bibfnamefont {S.}~\bibnamefont
  {{Chandrasekhar}}},\ }\bibfield  {title} {\enquote {\bibinfo {title}
  {{S}olutions of {T}wo {P}roblems in the {T}heory of {G}ravitational
  {R}adiation},}\ }\href {\doibase 10.1103/PhysRevLett.24.611} {\bibfield
  {journal} {\bibinfo  {journal} {Phys. Rev. Lett.}\ }\textbf {\bibinfo
  {volume} {24}},\ \bibinfo {pages} {611--615} (\bibinfo {year}
  {1970})}\BibitemShut {NoStop}%
\bibitem [{\citenamefont {{Friedman}}\ and\ \citenamefont
  {{Schutz}}(1978{\natexlab{a}})}]{FriedmanSchutz1978}%
  \BibitemOpen
  \bibfield  {author} {\bibinfo {author} {\bibfnamefont {J.~L.}\ \bibnamefont
  {{Friedman}}}\ and\ \bibinfo {author} {\bibfnamefont {B.~F.}\ \bibnamefont
  {{Schutz}}},\ }\bibfield  {title} {\enquote {\bibinfo {title} {{L}agrangian
  perturbation theory of nonrelativistic fluids},}\ }\href {\doibase
  10.1086/156098} {\bibfield  {journal} {\bibinfo  {journal} {Astrophys. J.}\
  }\textbf {\bibinfo {volume} {221}},\ \bibinfo {pages} {937--957} (\bibinfo
  {year} {1978}{\natexlab{a}})}\BibitemShut {NoStop}%
\bibitem [{\citenamefont {{Friedman}}\ and\ \citenamefont
  {{Schutz}}(1978{\natexlab{b}})}]{FriedmanSchutz1978b}%
  \BibitemOpen
  \bibfield  {author} {\bibinfo {author} {\bibfnamefont {J.~L.}\ \bibnamefont
  {{Friedman}}}\ and\ \bibinfo {author} {\bibfnamefont {B.~F.}\ \bibnamefont
  {{Schutz}}},\ }\bibfield  {title} {\enquote {\bibinfo {title} {{S}ecular
  instability of rotating {N}ewtonian stars},}\ }\href {\doibase
  10.1086/156143} {\bibfield  {journal} {\bibinfo  {journal} {Astrophys. J.}\
  }\textbf {\bibinfo {volume} {222}},\ \bibinfo {pages} {281--296} (\bibinfo
  {year} {1978}{\natexlab{b}})}\BibitemShut {NoStop}%
\bibitem [{\citenamefont {{Andersson}}(2003)}]{Andersson2003}%
  \BibitemOpen
  \bibfield  {author} {\bibinfo {author} {\bibfnamefont {N.}~\bibnamefont
  {{Andersson}}},\ }\bibfield  {title} {\enquote {\bibinfo {title} {{TOPICAL}
  {REVIEW}: {G}ravitational waves from instabilities in relativistic stars},}\
  }\href {\doibase 10.1088/0264-9381/20/7/201} {\bibfield  {journal} {\bibinfo
  {journal} {Classical Quantum Gravity}\ }\textbf {\bibinfo {volume} {20}},\
  \bibinfo {pages} {105} (\bibinfo {year} {2003})},\ \Eprint
  {http://arxiv.org/abs/astro-ph/0211057} {astro-ph/0211057} \BibitemShut
  {NoStop}%
\bibitem [{\citenamefont {{Pnigouras}}(2018)}]{Pnigouras2018}%
  \BibitemOpen
  \bibfield  {author} {\bibinfo {author} {\bibfnamefont {P.}~\bibnamefont
  {{Pnigouras}}},\ }\href {\doibase 10.1007/978-3-319-98258-8} {\emph {\bibinfo
  {title} {Saturation of the {$f$}-mode Instability in Neutron Stars}}},\
  Springer Theses\ (\bibinfo  {publisher} {Springer International Publishing},\
  \bibinfo {address} {Cham, Switzerland},\ \bibinfo {year} {2018})\BibitemShut
  {NoStop}%
\bibitem [{\citenamefont {{Papaloizou}}\ and\ \citenamefont
  {{Pringle}}(1978)}]{PapaloizouPringle1978}%
  \BibitemOpen
  \bibfield  {author} {\bibinfo {author} {\bibfnamefont {J.}~\bibnamefont
  {{Papaloizou}}}\ and\ \bibinfo {author} {\bibfnamefont {J.~E.}\ \bibnamefont
  {{Pringle}}},\ }\bibfield  {title} {\enquote {\bibinfo {title} {{N}on-radial
  oscillations of rotating stars and their relevance to the short-period
  oscillations of cataclysmic variables},}\ }\href
  {http://adsabs.harvard.edu/abs/1978MNRAS.182..423P} {\bibfield  {journal}
  {\bibinfo  {journal} {Mon. Not. R. Astron. Soc.}\ }\textbf {\bibinfo {volume}
  {182}},\ \bibinfo {pages} {423--442} (\bibinfo {year} {1978})}\BibitemShut
  {NoStop}%
\bibitem [{\citenamefont {{Andersson}}(1998)}]{Andersson1998}%
  \BibitemOpen
  \bibfield  {author} {\bibinfo {author} {\bibfnamefont {N.}~\bibnamefont
  {{Andersson}}},\ }\bibfield  {title} {\enquote {\bibinfo {title} {{A} {N}ew
  {C}lass of {U}nstable {M}odes of {R}otating {R}elativistic {S}tars},}\ }\href
  {\doibase 10.1086/305919} {\bibfield  {journal} {\bibinfo  {journal}
  {Astrophys. J.}\ }\textbf {\bibinfo {volume} {502}},\ \bibinfo {pages}
  {708--713} (\bibinfo {year} {1998})},\ \Eprint
  {http://arxiv.org/abs/gr-qc/9706075} {gr-qc/9706075} \BibitemShut {NoStop}%
\bibitem [{\citenamefont {{Friedman}}\ and\ \citenamefont
  {{Morsink}}(1998)}]{FriedmanMorsink1998}%
  \BibitemOpen
  \bibfield  {author} {\bibinfo {author} {\bibfnamefont {J.~L.}\ \bibnamefont
  {{Friedman}}}\ and\ \bibinfo {author} {\bibfnamefont {S.~M.}\ \bibnamefont
  {{Morsink}}},\ }\bibfield  {title} {\enquote {\bibinfo {title} {{A}xial
  {I}nstability of {R}otating {R}elativistic {S}tars},}\ }\href {\doibase
  10.1086/305920} {\bibfield  {journal} {\bibinfo  {journal} {Astrophys. J.}\
  }\textbf {\bibinfo {volume} {502}},\ \bibinfo {pages} {714--720} (\bibinfo
  {year} {1998})},\ \Eprint {http://arxiv.org/abs/gr-qc/9706073}
  {gr-qc/9706073} \BibitemShut {NoStop}%
\bibitem [{\citenamefont {{Lai}}\ and\ \citenamefont
  {{Shapiro}}(1995)}]{LaiShapiro1995}%
  \BibitemOpen
  \bibfield  {author} {\bibinfo {author} {\bibfnamefont {D.}~\bibnamefont
  {{Lai}}}\ and\ \bibinfo {author} {\bibfnamefont {S.~L.}\ \bibnamefont
  {{Shapiro}}},\ }\bibfield  {title} {\enquote {\bibinfo {title}
  {{G}ravitational radiation from rapidly rotating nascent neutron stars},}\
  }\href {\doibase 10.1086/175438} {\bibfield  {journal} {\bibinfo  {journal}
  {Astrophys. J.}\ }\textbf {\bibinfo {volume} {442}},\ \bibinfo {pages}
  {259--272} (\bibinfo {year} {1995})},\ \Eprint
  {http://arxiv.org/abs/astro-ph/9408053} {astro-ph/9408053} \BibitemShut
  {NoStop}%
\bibitem [{\citenamefont {{Owen}}\ \emph {et~al.}(1998)\citenamefont {{Owen}},
  \citenamefont {{Lindblom}}, \citenamefont {{Cutler}}, \citenamefont
  {{Schutz}}, \citenamefont {{Vecchio}},\ and\ \citenamefont
  {{Andersson}}}]{OwenEtAl1998}%
  \BibitemOpen
  \bibfield  {author} {\bibinfo {author} {\bibfnamefont {B.~J.}\ \bibnamefont
  {{Owen}}}, \bibinfo {author} {\bibfnamefont {L.}~\bibnamefont {{Lindblom}}},
  \bibinfo {author} {\bibfnamefont {C.}~\bibnamefont {{Cutler}}}, \bibinfo
  {author} {\bibfnamefont {B.~F.}\ \bibnamefont {{Schutz}}}, \bibinfo {author}
  {\bibfnamefont {A.}~\bibnamefont {{Vecchio}}}, \ and\ \bibinfo {author}
  {\bibfnamefont {N.}~\bibnamefont {{Andersson}}},\ }\bibfield  {title}
  {\enquote {\bibinfo {title} {{G}ravitational waves from hot young rapidly
  rotating neutron stars},}\ }\href {\doibase 10.1103/PhysRevD.58.084020}
  {\bibfield  {journal} {\bibinfo  {journal} {Phys. Rev. D}\ }\textbf {\bibinfo
  {volume} {58}},\ \bibinfo {eid} {084020} (\bibinfo {year} {1998})},\ \Eprint
  {http://arxiv.org/abs/gr-qc/9804044} {gr-qc/9804044} \BibitemShut {NoStop}%
\bibitem [{\citenamefont {{Doneva}}\ \emph {et~al.}(2013)\citenamefont
  {{Doneva}}, \citenamefont {{Gaertig}}, \citenamefont {{Kokkotas}},\ and\
  \citenamefont {{Kr{\"u}ger}}}]{DonevaEtAl2013}%
  \BibitemOpen
  \bibfield  {author} {\bibinfo {author} {\bibfnamefont {D.~D.}\ \bibnamefont
  {{Doneva}}}, \bibinfo {author} {\bibfnamefont {E.}~\bibnamefont {{Gaertig}}},
  \bibinfo {author} {\bibfnamefont {K.~D.}\ \bibnamefont {{Kokkotas}}}, \ and\
  \bibinfo {author} {\bibfnamefont {C.}~\bibnamefont {{Kr{\"u}ger}}},\
  }\bibfield  {title} {\enquote {\bibinfo {title} {{G}ravitational wave
  asteroseismology of fast rotating neutron stars with realistic equations of
  state},}\ }\href {\doibase 10.1103/PhysRevD.88.044052} {\bibfield  {journal}
  {\bibinfo  {journal} {Phys. Rev. D}\ }\textbf {\bibinfo {volume} {88}},\
  \bibinfo {eid} {044052} (\bibinfo {year} {2013})},\ \Eprint
  {http://arxiv.org/abs/1305.7197} {arXiv:1305.7197 [astro-ph.SR]} \BibitemShut
  {NoStop}%
\bibitem [{\citenamefont {{Alford}}\ and\ \citenamefont
  {{Schwenzer}}(2014)}]{AlfordSchwenzer2014}%
  \BibitemOpen
  \bibfield  {author} {\bibinfo {author} {\bibfnamefont {M.~G.}\ \bibnamefont
  {{Alford}}}\ and\ \bibinfo {author} {\bibfnamefont {K.}~\bibnamefont
  {{Schwenzer}}},\ }\bibfield  {title} {\enquote {\bibinfo {title} {{W}hat the
  {T}iming of {M}illisecond {P}ulsars {C}an {T}each us about {T}heir
  {I}nterior},}\ }\href {\doibase 10.1103/PhysRevLett.113.251102} {\bibfield
  {journal} {\bibinfo  {journal} {Phys. Rev. Lett.}\ }\textbf {\bibinfo
  {volume} {113}},\ \bibinfo {eid} {251102} (\bibinfo {year} {2014})},\ \Eprint
  {http://arxiv.org/abs/1310.3524} {arXiv:1310.3524 [astro-ph.HE]} \BibitemShut
  {NoStop}%
\bibitem [{\citenamefont {{Doneva}}\ \emph {et~al.}(2015)\citenamefont
  {{Doneva}}, \citenamefont {{Kokkotas}},\ and\ \citenamefont
  {{Pnigouras}}}]{DonevaEtAl2015}%
  \BibitemOpen
  \bibfield  {author} {\bibinfo {author} {\bibfnamefont {D.~D.}\ \bibnamefont
  {{Doneva}}}, \bibinfo {author} {\bibfnamefont {K.~D.}\ \bibnamefont
  {{Kokkotas}}}, \ and\ \bibinfo {author} {\bibfnamefont {P.}~\bibnamefont
  {{Pnigouras}}},\ }\bibfield  {title} {\enquote {\bibinfo {title}
  {{G}ravitational wave afterglow in binary neutron star mergers},}\ }\href
  {\doibase 10.1103/PhysRevD.92.104040} {\bibfield  {journal} {\bibinfo
  {journal} {Phys. Rev. D}\ }\textbf {\bibinfo {volume} {92}},\ \bibinfo {eid}
  {104040} (\bibinfo {year} {2015})},\ \Eprint
  {http://arxiv.org/abs/1510.00673} {arXiv:1510.00673 [gr-qc]} \BibitemShut
  {NoStop}%
\bibitem [{\citenamefont {{Levin}}(1999)}]{Levin1999}%
  \BibitemOpen
  \bibfield  {author} {\bibinfo {author} {\bibfnamefont {Y.}~\bibnamefont
  {{Levin}}},\ }\bibfield  {title} {\enquote {\bibinfo {title} {{R}unaway
  {H}eating by {R}-{M}odes of {N}eutron {S}tars in {L}ow-{M}ass {X}-{R}ay
  {B}inaries},}\ }\href {\doibase 10.1086/307196} {\bibfield  {journal}
  {\bibinfo  {journal} {Astrophys. J.}\ }\textbf {\bibinfo {volume} {517}},\
  \bibinfo {pages} {328--333} (\bibinfo {year} {1999})},\ \Eprint
  {http://arxiv.org/abs/astro-ph/9810471} {astro-ph/9810471} \BibitemShut
  {NoStop}%
\bibitem [{\citenamefont {{Andersson}}\ \emph {et~al.}(2000)\citenamefont
  {{Andersson}}, \citenamefont {{Jones}}, \citenamefont {{Kokkotas}},\ and\
  \citenamefont {{Stergioulas}}}]{AnderssonEtAl2000}%
  \BibitemOpen
  \bibfield  {author} {\bibinfo {author} {\bibfnamefont {N.}~\bibnamefont
  {{Andersson}}}, \bibinfo {author} {\bibfnamefont {D.~I.}\ \bibnamefont
  {{Jones}}}, \bibinfo {author} {\bibfnamefont {K.~D.}\ \bibnamefont
  {{Kokkotas}}}, \ and\ \bibinfo {author} {\bibfnamefont {N.}~\bibnamefont
  {{Stergioulas}}},\ }\bibfield  {title} {\enquote {\bibinfo {title}
  {{R}-{M}ode {R}unaway and {R}apidly {R}otating {N}eutron {S}tars},}\ }\href
  {\doibase 10.1086/312643} {\bibfield  {journal} {\bibinfo  {journal}
  {Astrophys. J.}\ }\textbf {\bibinfo {volume} {534}},\ \bibinfo {pages}
  {L75--L78} (\bibinfo {year} {2000})},\ \Eprint
  {http://arxiv.org/abs/astro-ph/0002114} {astro-ph/0002114} \BibitemShut
  {NoStop}%
\bibitem [{\citenamefont {{Passamonti}}\ \emph {et~al.}(2013)\citenamefont
  {{Passamonti}}, \citenamefont {{Gaertig}}, \citenamefont {{Kokkotas}},\ and\
  \citenamefont {{Doneva}}}]{PassamontiEtAl2013}%
  \BibitemOpen
  \bibfield  {author} {\bibinfo {author} {\bibfnamefont {A.}~\bibnamefont
  {{Passamonti}}}, \bibinfo {author} {\bibfnamefont {E.}~\bibnamefont
  {{Gaertig}}}, \bibinfo {author} {\bibfnamefont {K.~D.}\ \bibnamefont
  {{Kokkotas}}}, \ and\ \bibinfo {author} {\bibfnamefont {D.}~\bibnamefont
  {{Doneva}}},\ }\bibfield  {title} {\enquote {\bibinfo {title} {{E}volution of
  the f-mode instability in neutron stars and gravitational wave
  detectability},}\ }\href {\doibase 10.1103/PhysRevD.87.084010} {\bibfield
  {journal} {\bibinfo  {journal} {Phys. Rev. D}\ }\textbf {\bibinfo {volume}
  {87}},\ \bibinfo {eid} {084010} (\bibinfo {year} {2013})},\ \Eprint
  {http://arxiv.org/abs/1209.5308} {arXiv:1209.5308 [astro-ph.SR]} \BibitemShut
  {NoStop}%
\bibitem [{\citenamefont {{Bondarescu}}\ \emph {et~al.}(2007)\citenamefont
  {{Bondarescu}}, \citenamefont {{Teukolsky}},\ and\ \citenamefont
  {{Wasserman}}}]{BondarescuEtAl2007}%
  \BibitemOpen
  \bibfield  {author} {\bibinfo {author} {\bibfnamefont {R.}~\bibnamefont
  {{Bondarescu}}}, \bibinfo {author} {\bibfnamefont {S.~A.}\ \bibnamefont
  {{Teukolsky}}}, \ and\ \bibinfo {author} {\bibfnamefont {I.}~\bibnamefont
  {{Wasserman}}},\ }\bibfield  {title} {\enquote {\bibinfo {title} {{S}pin
  evolution of accreting neutron stars: {N}onlinear development of the r-mode
  instability},}\ }\href {\doibase 10.1103/PhysRevD.76.064019} {\bibfield
  {journal} {\bibinfo  {journal} {Phys. Rev. D}\ }\textbf {\bibinfo {volume}
  {76}},\ \bibinfo {eid} {064019} (\bibinfo {year} {2007})},\ \Eprint
  {http://arxiv.org/abs/0704.0799} {arXiv:0704.0799} \BibitemShut {NoStop}%
\bibitem [{\citenamefont {{Bondarescu}}\ \emph {et~al.}(2009)\citenamefont
  {{Bondarescu}}, \citenamefont {{Teukolsky}},\ and\ \citenamefont
  {{Wasserman}}}]{BondarescuEtAl2009}%
  \BibitemOpen
  \bibfield  {author} {\bibinfo {author} {\bibfnamefont {R.}~\bibnamefont
  {{Bondarescu}}}, \bibinfo {author} {\bibfnamefont {S.~A.}\ \bibnamefont
  {{Teukolsky}}}, \ and\ \bibinfo {author} {\bibfnamefont {I.}~\bibnamefont
  {{Wasserman}}},\ }\bibfield  {title} {\enquote {\bibinfo {title} {{S}pinning
  down newborn neutron stars: {N}onlinear development of the r-mode
  instability},}\ }\href {\doibase 10.1103/PhysRevD.79.104003} {\bibfield
  {journal} {\bibinfo  {journal} {Phys. Rev. D}\ }\textbf {\bibinfo {volume}
  {79}},\ \bibinfo {eid} {104003} (\bibinfo {year} {2009})},\ \Eprint
  {http://arxiv.org/abs/0809.3448} {arXiv:0809.3448} \BibitemShut {NoStop}%
\bibitem [{\citenamefont {{Friedman}}(1983)}]{Friedman1983}%
  \BibitemOpen
  \bibfield  {author} {\bibinfo {author} {\bibfnamefont {J.~L.}\ \bibnamefont
  {{Friedman}}},\ }\bibfield  {title} {\enquote {\bibinfo {title} {{U}pper
  limit on the frequency of pulsars},}\ }\href {\doibase
  10.1103/PhysRevLett.51.11} {\bibfield  {journal} {\bibinfo  {journal} {Phys.
  Rev. Lett.}\ }\textbf {\bibinfo {volume} {51}},\ \bibinfo {pages} {11--14}
  (\bibinfo {year} {1983})}\BibitemShut {NoStop}%
\bibitem [{\citenamefont {{Andersson}}\ \emph
  {et~al.}(1999{\natexlab{a}})\citenamefont {{Andersson}}, \citenamefont
  {{Kokkotas}},\ and\ \citenamefont {{Schutz}}}]{AnderssonEtAl1999}%
  \BibitemOpen
  \bibfield  {author} {\bibinfo {author} {\bibfnamefont {N.}~\bibnamefont
  {{Andersson}}}, \bibinfo {author} {\bibfnamefont {K.}~\bibnamefont
  {{Kokkotas}}}, \ and\ \bibinfo {author} {\bibfnamefont {B.~F.}\ \bibnamefont
  {{Schutz}}},\ }\bibfield  {title} {\enquote {\bibinfo {title}
  {{G}ravitational {R}adiation {L}imit on the {S}pin of {Y}oung {N}eutron
  {S}tars},}\ }\href {\doibase 10.1086/306625} {\bibfield  {journal} {\bibinfo
  {journal} {Astrophys. J.}\ }\textbf {\bibinfo {volume} {510}},\ \bibinfo
  {pages} {846--853} (\bibinfo {year} {1999}{\natexlab{a}})},\ \Eprint
  {http://arxiv.org/abs/astro-ph/9805225} {astro-ph/9805225} \BibitemShut
  {NoStop}%
\bibitem [{\citenamefont {{Andersson}}\ \emph
  {et~al.}(1999{\natexlab{b}})\citenamefont {{Andersson}}, \citenamefont
  {{Kokkotas}},\ and\ \citenamefont {{Stergioulas}}}]{AnderssonEtAl1999b}%
  \BibitemOpen
  \bibfield  {author} {\bibinfo {author} {\bibfnamefont {N.}~\bibnamefont
  {{Andersson}}}, \bibinfo {author} {\bibfnamefont {K.~D.}\ \bibnamefont
  {{Kokkotas}}}, \ and\ \bibinfo {author} {\bibfnamefont {N.}~\bibnamefont
  {{Stergioulas}}},\ }\bibfield  {title} {\enquote {\bibinfo {title} {{O}n the
  {R}elevance of the {R}-{M}ode {I}nstability for {A}ccreting {N}eutron {S}tars
  and {W}hite {D}warfs},}\ }\href {\doibase 10.1086/307082} {\bibfield
  {journal} {\bibinfo  {journal} {Astrophys. J.}\ }\textbf {\bibinfo {volume}
  {516}},\ \bibinfo {pages} {307--314} (\bibinfo {year}
  {1999}{\natexlab{b}})},\ \Eprint {http://arxiv.org/abs/astro-ph/9806089}
  {astro-ph/9806089} \BibitemShut {NoStop}%
\bibitem [{\citenamefont {Abbott}\ \emph
  {et~al.}(2017{\natexlab{b}})\citenamefont {Abbott} \emph
  {et~al.}}]{AbbottEtAl2017t}%
  \BibitemOpen
  \bibfield  {author} {\bibinfo {author} {\bibfnamefont {B.~P.}\ \bibnamefont
  {Abbott}} \emph {et~al.},\ }\bibfield  {title} {\enquote {\bibinfo {title}
  {Search for Post-merger Gravitational Waves from the Remnant of the Binary
  Neutron Star Merger GW170817},}\ }\href {\doibase 10.3847/2041-8213/aa9a35}
  {\bibfield  {journal} {\bibinfo  {journal} {Astrophys. J.}\ }\textbf
  {\bibinfo {volume} {851}},\ \bibinfo {eid} {L16} (\bibinfo {year}
  {2017}{\natexlab{b}})},\ \Eprint {http://arxiv.org/abs/1710.09320}
  {arXiv:1710.09320 [astro-ph.HE]} \BibitemShut {NoStop}%
\bibitem [{\citenamefont {Abbott}\ \emph
  {et~al.}(2019{\natexlab{a}})\citenamefont {Abbott} \emph
  {et~al.}}]{AbbottEtAl2019}%
  \BibitemOpen
  \bibfield  {author} {\bibinfo {author} {\bibfnamefont {B.~P.}\ \bibnamefont
  {Abbott}} \emph {et~al.},\ }\bibfield  {title} {\enquote {\bibinfo {title}
  {Search for Gravitational Waves from a Long-lived Remnant of the Binary
  Neutron Star Merger GW170817},}\ }\href {\doibase 10.3847/1538-4357/ab0f3d}
  {\bibfield  {journal} {\bibinfo  {journal} {Astrophys. J.}\ }\textbf
  {\bibinfo {volume} {875}},\ \bibinfo {eid} {160} (\bibinfo {year}
  {2019}{\natexlab{a}})},\ \Eprint {http://arxiv.org/abs/1810.02581}
  {arXiv:1810.02581 [gr-qc]} \BibitemShut {NoStop}%
\bibitem [{\citenamefont {Abbott}\ \emph
  {et~al.}(2019{\natexlab{b}})\citenamefont {Abbott} \emph
  {et~al.}}]{AbbottEtAl2019b}%
  \BibitemOpen
  \bibfield  {author} {\bibinfo {author} {\bibfnamefont {B.~P.}\ \bibnamefont
  {Abbott}} \emph {et~al.},\ }\bibfield  {title} {\enquote {\bibinfo {title}
  {Searches for Continuous Gravitational Waves from 15 Supernova Remnants and
  Fomalhaut b with Advanced LIGO},}\ }\href {\doibase 10.3847/1538-4357/ab113b}
  {\bibfield  {journal} {\bibinfo  {journal} {Astrophys. J.}\ }\textbf
  {\bibinfo {volume} {875}},\ \bibinfo {eid} {122} (\bibinfo {year}
  {2019}{\natexlab{b}})},\ \Eprint {http://arxiv.org/abs/1812.11656}
  {arXiv:1812.11656 [gr-qc]} \BibitemShut {NoStop}%
\bibitem [{\citenamefont {Abbott}\ \emph
  {et~al.}(2019{\natexlab{c}})\citenamefont {Abbott} \emph
  {et~al.}}]{AbbottEtAl2019c}%
  \BibitemOpen
  \bibfield  {author} {\bibinfo {author} {\bibfnamefont {B.~P.}\ \bibnamefont
  {Abbott}} \emph {et~al.},\ }\bibfield  {title} {\enquote {\bibinfo {title}
  {All-sky search for long-duration gravitational-wave transients in the second
  Advanced LIGO observing run},}\ }\href {\doibase 10.1103/PhysRevD.99.104033}
  {\bibfield  {journal} {\bibinfo  {journal} {Phys. Rev. D}\ }\textbf {\bibinfo
  {volume} {99}},\ \bibinfo {eid} {104033} (\bibinfo {year}
  {2019}{\natexlab{c}})},\ \Eprint {http://arxiv.org/abs/1903.12015}
  {arXiv:1903.12015 [gr-qc]} \BibitemShut {NoStop}%
\bibitem [{\citenamefont {Mashhoon}(1973)}]{Mashhoon1973}%
  \BibitemOpen
  \bibfield  {author} {\bibinfo {author} {\bibfnamefont {B.}~\bibnamefont
  {Mashhoon}},\ }\bibfield  {title} {\enquote {\bibinfo {title} {Tidal
  Gravitational Radiation},}\ }\href {\doibase 10.1086/152397} {\bibfield
  {journal} {\bibinfo  {journal} {Astrophys. J.}\ }\textbf {\bibinfo {volume}
  {185}},\ \bibinfo {pages} {83--86} (\bibinfo {year} {1973})}\BibitemShut
  {NoStop}%
\bibitem [{\citenamefont {Mashhoon}(1975)}]{Mashhoon1975}%
  \BibitemOpen
  \bibfield  {author} {\bibinfo {author} {\bibfnamefont {B.}~\bibnamefont
  {Mashhoon}},\ }\bibfield  {title} {\enquote {\bibinfo {title} {On tidal
  phenomena in a strong gravitational field},}\ }\href {\doibase
  10.1086/153560} {\bibfield  {journal} {\bibinfo  {journal} {Astrophys. J.}\
  }\textbf {\bibinfo {volume} {197}},\ \bibinfo {pages} {705--716} (\bibinfo
  {year} {1975})}\BibitemShut {NoStop}%
\bibitem [{\citenamefont {Mashhoon}(1977)}]{Mashhoon1977}%
  \BibitemOpen
  \bibfield  {author} {\bibinfo {author} {\bibfnamefont {B.}~\bibnamefont
  {Mashhoon}},\ }\bibfield  {title} {\enquote {\bibinfo {title} {Tidal
  radiation},}\ }\href {\doibase 10.1086/155500} {\bibfield  {journal}
  {\bibinfo  {journal} {Astrophys. J.}\ }\textbf {\bibinfo {volume} {216}},\
  \bibinfo {pages} {591--609} (\bibinfo {year} {1977})}\BibitemShut {NoStop}%
\bibitem [{\citenamefont {Turner}(1977)}]{Turner1977}%
  \BibitemOpen
  \bibfield  {author} {\bibinfo {author} {\bibfnamefont {M.}~\bibnamefont
  {Turner}},\ }\bibfield  {title} {\enquote {\bibinfo {title} {Tidal generation
  of gravitational waves from orbiting Newtonian stars. I - General
  formalism},}\ }\href {\doibase 10.1086/155536} {\bibfield  {journal}
  {\bibinfo  {journal} {Astrophys. J.}\ }\textbf {\bibinfo {volume} {216}},\
  \bibinfo {pages} {914--929} (\bibinfo {year} {1977})}\BibitemShut {NoStop}%
\bibitem [{\citenamefont {Will}(1983)}]{Will1983}%
  \BibitemOpen
  \bibfield  {author} {\bibinfo {author} {\bibfnamefont {C.~M.}\ \bibnamefont
  {Will}},\ }\bibfield  {title} {\enquote {\bibinfo {title} {Tidal
  gravitational radiation from homogeneous stars},}\ }\href {\doibase
  10.1086/161499} {\bibfield  {journal} {\bibinfo  {journal} {Astrophys. J.}\
  }\textbf {\bibinfo {volume} {274}},\ \bibinfo {pages} {858--874} (\bibinfo
  {year} {1983})}\BibitemShut {NoStop}%
\bibitem [{\citenamefont {Ho}\ and\ \citenamefont {Lai}(1999)}]{HoLai1999}%
  \BibitemOpen
  \bibfield  {author} {\bibinfo {author} {\bibfnamefont {W.~C.~G.}\
  \bibnamefont {Ho}}\ and\ \bibinfo {author} {\bibfnamefont {D.}~\bibnamefont
  {Lai}},\ }\bibfield  {title} {\enquote {\bibinfo {title} {Resonant tidal
  excitations of rotating neutron stars in coalescing binaries},}\ }\href
  {\doibase 10.1046/j.1365-8711.1999.02703.x} {\bibfield  {journal} {\bibinfo
  {journal} {Mon. Not. R. Astron. Soc.}\ }\textbf {\bibinfo {volume} {308}},\
  \bibinfo {pages} {153--166} (\bibinfo {year} {1999})},\ \Eprint
  {http://arxiv.org/abs/astro-ph/9812116} {astro-ph/9812116} \BibitemShut
  {NoStop}%
\bibitem [{\citenamefont {{Reisenegger}}\ and\ \citenamefont
  {{Goldreich}}(1992)}]{ReiseneggerGoldreich1992}%
  \BibitemOpen
  \bibfield  {author} {\bibinfo {author} {\bibfnamefont {A.}~\bibnamefont
  {{Reisenegger}}}\ and\ \bibinfo {author} {\bibfnamefont {P.}~\bibnamefont
  {{Goldreich}}},\ }\bibfield  {title} {\enquote {\bibinfo {title} {{A} new
  class of g-modes in neutron stars},}\ }\href {\doibase 10.1086/171645}
  {\bibfield  {journal} {\bibinfo  {journal} {Astrophys. J.}\ }\textbf
  {\bibinfo {volume} {395}},\ \bibinfo {pages} {240--249} (\bibinfo {year}
  {1992})}\BibitemShut {NoStop}%
\bibitem [{\citenamefont {{Unno}}\ \emph {et~al.}(1989)\citenamefont {{Unno}},
  \citenamefont {{Osaki}}, \citenamefont {{Ando}}, \citenamefont {{Saio}},\
  and\ \citenamefont {{Shibahashi}}}]{UnnoEtAl1989}%
  \BibitemOpen
  \bibfield  {author} {\bibinfo {author} {\bibfnamefont {W.}~\bibnamefont
  {{Unno}}}, \bibinfo {author} {\bibfnamefont {Y.}~\bibnamefont {{Osaki}}},
  \bibinfo {author} {\bibfnamefont {H.}~\bibnamefont {{Ando}}}, \bibinfo
  {author} {\bibfnamefont {H.}~\bibnamefont {{Saio}}}, \ and\ \bibinfo {author}
  {\bibfnamefont {H.}~\bibnamefont {{Shibahashi}}},\ }\href
  {http://adsabs.harvard.edu/abs/1989nos..book.....U} {\emph {\bibinfo {title}
  {{N}onradial {O}scillations of {S}tars}}},\ \bibinfo {edition} {2nd}\ ed.\
  (\bibinfo  {publisher} {University of Tokyo Press},\ \bibinfo {address}
  {Tokyo},\ \bibinfo {year} {1989})\BibitemShut {NoStop}%
\bibitem [{\citenamefont {{Kr{\"u}ger}}\ \emph {et~al.}(2015)\citenamefont
  {{Kr{\"u}ger}}, \citenamefont {{Ho}},\ and\ \citenamefont
  {{Andersson}}}]{KruegerEtAl2015}%
  \BibitemOpen
  \bibfield  {author} {\bibinfo {author} {\bibfnamefont {C.~J.}\ \bibnamefont
  {{Kr{\"u}ger}}}, \bibinfo {author} {\bibfnamefont {W.~C.~G.}\ \bibnamefont
  {{Ho}}}, \ and\ \bibinfo {author} {\bibfnamefont {N.}~\bibnamefont
  {{Andersson}}},\ }\bibfield  {title} {\enquote {\bibinfo {title}
  {{S}eismology of adolescent neutron stars: {A}ccounting for thermal effects
  and crust elasticity},}\ }\href {\doibase 10.1103/PhysRevD.92.063009}
  {\bibfield  {journal} {\bibinfo  {journal} {Phys. Rev. D}\ }\textbf {\bibinfo
  {volume} {92}},\ \bibinfo {eid} {063009} (\bibinfo {year} {2015})},\ \Eprint
  {http://arxiv.org/abs/1402.5656} {arXiv:1402.5656 [gr-qc]} \BibitemShut
  {NoStop}%
\bibitem [{\citenamefont {{Schenk}}\ \emph {et~al.}(2001)\citenamefont
  {{Schenk}}, \citenamefont {{Arras}}, \citenamefont {{Flanagan}},
  \citenamefont {{Teukolsky}},\ and\ \citenamefont
  {{Wasserman}}}]{SchenkEtAl2001}%
  \BibitemOpen
  \bibfield  {author} {\bibinfo {author} {\bibfnamefont {A.~K.}\ \bibnamefont
  {{Schenk}}}, \bibinfo {author} {\bibfnamefont {P.}~\bibnamefont {{Arras}}},
  \bibinfo {author} {\bibfnamefont {{\'E}.~{\'E}.}\ \bibnamefont {{Flanagan}}},
  \bibinfo {author} {\bibfnamefont {S.~A.}\ \bibnamefont {{Teukolsky}}}, \ and\
  \bibinfo {author} {\bibfnamefont {I.}~\bibnamefont {{Wasserman}}},\
  }\bibfield  {title} {\enquote {\bibinfo {title} {{N}onlinear mode coupling in
  rotating stars and the r-mode instability in neutron stars},}\ }\href
  {\doibase 10.1103/PhysRevD.65.024001} {\bibfield  {journal} {\bibinfo
  {journal} {Phys. Rev. D}\ }\textbf {\bibinfo {volume} {65}},\ \bibinfo {eid}
  {024001} (\bibinfo {year} {2001})},\ \Eprint
  {http://arxiv.org/abs/gr-qc/0101092} {gr-qc/0101092} \BibitemShut {NoStop}%
\bibitem [{\citenamefont {{Thorne}}(1969)}]{Thorne1969}%
  \BibitemOpen
  \bibfield  {author} {\bibinfo {author} {\bibfnamefont {K.~S.}\ \bibnamefont
  {{Thorne}}},\ }\bibfield  {title} {\enquote {\bibinfo {title} {{N}onradial
  {P}ulsation of {G}eneral-{R}elativistic {S}tellar {M}odels. {IV}. {T}he
  {W}eak-field {L}imit},}\ }\href {\doibase 10.1086/150259} {\bibfield
  {journal} {\bibinfo  {journal} {Astrophys. J.}\ }\textbf {\bibinfo {volume}
  {158}},\ \bibinfo {pages} {997} (\bibinfo {year} {1969})}\BibitemShut
  {NoStop}%
\bibitem [{\citenamefont {{Ipser}}\ and\ \citenamefont
  {{Lindblom}}(1991)}]{IpserLindblom1991}%
  \BibitemOpen
  \bibfield  {author} {\bibinfo {author} {\bibfnamefont {J.~R.}\ \bibnamefont
  {{Ipser}}}\ and\ \bibinfo {author} {\bibfnamefont {L.}~\bibnamefont
  {{Lindblom}}},\ }\bibfield  {title} {\enquote {\bibinfo {title} {{T}he
  oscillations of rapidly rotating {N}ewtonian stellar models. {II}.
  {D}issipative effects},}\ }\href {\doibase 10.1086/170039} {\bibfield
  {journal} {\bibinfo  {journal} {Astrophys. J.}\ }\textbf {\bibinfo {volume}
  {373}},\ \bibinfo {pages} {213--221} (\bibinfo {year} {1991})}\BibitemShut
  {NoStop}%
\bibitem [{\citenamefont {{Thorne}}(1980)}]{Thorne1980}%
  \BibitemOpen
  \bibfield  {author} {\bibinfo {author} {\bibfnamefont {K.~S.}\ \bibnamefont
  {{Thorne}}},\ }\bibfield  {title} {\enquote {\bibinfo {title} {{M}ultipole
  expansions of gravitational radiation},}\ }\href {\doibase
  10.1103/RevModPhys.52.299} {\bibfield  {journal} {\bibinfo  {journal} {Rev.
  Mod. Phys.}\ }\textbf {\bibinfo {volume} {52}},\ \bibinfo {pages} {299--340}
  (\bibinfo {year} {1980})}\BibitemShut {NoStop}%
\bibitem [{\citenamefont {{Zahn}}(1966)}]{Zahn1966}%
  \BibitemOpen
  \bibfield  {author} {\bibinfo {author} {\bibfnamefont {J.~P.}\ \bibnamefont
  {{Zahn}}},\ }\bibfield  {title} {\enquote {\bibinfo {title} {{Les mar{\'e}es
  dans une {\'e}toile double serr{\'e}e}},}\ }\href
  {http://adsabs.harvard.edu/abs/1966AnAp...29..313Z} {\bibfield  {journal}
  {\bibinfo  {journal} {Ann. Astrophys.}\ }\textbf {\bibinfo {volume} {29}},\
  \bibinfo {pages} {313} (\bibinfo {year} {1966})}\BibitemShut {NoStop}%
\bibitem [{\citenamefont {Ogilvie}(2014)}]{Ogilvie2014}%
  \BibitemOpen
  \bibfield  {author} {\bibinfo {author} {\bibfnamefont {G.~I.}\ \bibnamefont
  {Ogilvie}},\ }\bibfield  {title} {\enquote {\bibinfo {title} {Tidal
  Dissipation in Stars and Giant Planets},}\ }\href {\doibase
  10.1146/annurev-astro-081913-035941} {\bibfield  {journal} {\bibinfo
  {journal} {Annu. Rev. Astron. Astrophys.}\ }\textbf {\bibinfo {volume}
  {52}},\ \bibinfo {pages} {171--210} (\bibinfo {year} {2014})},\ \Eprint
  {http://arxiv.org/abs/1406.2207} {arXiv:1406.2207 [astro-ph.SR]} \BibitemShut
  {NoStop}%
\bibitem [{\citenamefont {Ogilvie}\ and\ \citenamefont
  {Lin}(2004)}]{OgilvieLin2004}%
  \BibitemOpen
  \bibfield  {author} {\bibinfo {author} {\bibfnamefont {G.~I.}\ \bibnamefont
  {Ogilvie}}\ and\ \bibinfo {author} {\bibfnamefont {D.~N.~C.}\ \bibnamefont
  {Lin}},\ }\bibfield  {title} {\enquote {\bibinfo {title} {Tidal Dissipation
  in Rotating Giant Planets},}\ }\href {\doibase 10.1086/421454} {\bibfield
  {journal} {\bibinfo  {journal} {Astrophys. J.}\ }\textbf {\bibinfo {volume}
  {610}},\ \bibinfo {pages} {477--509} (\bibinfo {year} {2004})},\ \Eprint
  {http://arxiv.org/abs/astro-ph/0310218} {astro-ph/0310218} \BibitemShut
  {NoStop}%
\bibitem [{\citenamefont {Press}\ and\ \citenamefont
  {Teukolsky}(1977)}]{PressTeukolsky1977}%
  \BibitemOpen
  \bibfield  {author} {\bibinfo {author} {\bibfnamefont {W.~H.}\ \bibnamefont
  {Press}}\ and\ \bibinfo {author} {\bibfnamefont {S.~A.}\ \bibnamefont
  {Teukolsky}},\ }\bibfield  {title} {\enquote {\bibinfo {title} {On formation
  of close binaries by two-body tidal capture},}\ }\href {\doibase
  10.1086/155143} {\bibfield  {journal} {\bibinfo  {journal} {Astrophys. J.}\
  }\textbf {\bibinfo {volume} {213}},\ \bibinfo {pages} {183--192} (\bibinfo
  {year} {1977})}\BibitemShut {NoStop}%
\bibitem [{\citenamefont {Steinborn}\ and\ \citenamefont
  {Ruedenberg}(1973)}]{SteinbornRuedenberg1973}%
  \BibitemOpen
  \bibfield  {author} {\bibinfo {author} {\bibfnamefont {E.~O.}\ \bibnamefont
  {Steinborn}}\ and\ \bibinfo {author} {\bibfnamefont {K.}~\bibnamefont
  {Ruedenberg}},\ }\bibfield  {title} {\enquote {\bibinfo {title} {Rotation and
  Translation of Regular and Irregular Solid Spherical Harmonics},}\ }\href
  {\doibase 10.1016/S0065-3276(08)60558-4} {\bibfield  {journal} {\bibinfo
  {journal} {Adv. Quantum Chem.}\ }\textbf {\bibinfo {volume} {7}},\ \bibinfo
  {pages} {1--81} (\bibinfo {year} {1973})}\BibitemShut {NoStop}%
\bibitem [{\citenamefont {{Shapiro}}\ and\ \citenamefont
  {{Teukolsky}}(1983)}]{ShapiroTeukolsky1983}%
  \BibitemOpen
  \bibfield  {author} {\bibinfo {author} {\bibfnamefont {S.~L.}\ \bibnamefont
  {{Shapiro}}}\ and\ \bibinfo {author} {\bibfnamefont {S.~A.}\ \bibnamefont
  {{Teukolsky}}},\ }\href {http://adsabs.harvard.edu/abs/1983bhwd.book.....S}
  {\emph {\bibinfo {title} {{B}lack holes, white dwarfs, and neutron stars:
  {T}he physics of compact objects}}}\ (\bibinfo  {publisher} {John Wiley \&
  Sons},\ \bibinfo {address} {New York},\ \bibinfo {year} {1983})\BibitemShut
  {NoStop}%
\bibitem [{\citenamefont {{Maggiore}}(2008)}]{Maggiore2008}%
  \BibitemOpen
  \bibfield  {author} {\bibinfo {author} {\bibfnamefont {Michele}\ \bibnamefont
  {{Maggiore}}},\ }\href
  {http://www.oxfordscholarship.com/view/10.1093/acprof:oso/9780198570745.001.0001/acprof-9780198570745}
  {\emph {\bibinfo {title} {{G}ravitational {W}aves: {T}heory and
  {E}xperiments}}},\ Vol.~\bibinfo {volume} {1}\ (\bibinfo  {publisher} {Oxford
  University Press},\ \bibinfo {address} {Oxford, England},\ \bibinfo {year}
  {2008})\BibitemShut {NoStop}%
\bibitem [{\citenamefont {Yip}\ and\ \citenamefont
  {Leung}(2017)}]{YipLeung2017}%
  \BibitemOpen
  \bibfield  {author} {\bibinfo {author} {\bibfnamefont {K.~L.~S.}\
  \bibnamefont {Yip}}\ and\ \bibinfo {author} {\bibfnamefont {P.~T.}\
  \bibnamefont {Leung}},\ }\bibfield  {title} {\enquote {\bibinfo {title}
  {Tidal Love numbers and moment-Love relations of polytropic stars},}\ }\href
  {\doibase 10.1093/mnras/stx2363} {\bibfield  {journal} {\bibinfo  {journal}
  {Mon. Not. R. Astron. Soc.}\ }\textbf {\bibinfo {volume} {472}},\ \bibinfo
  {pages} {4965--4981} (\bibinfo {year} {2017})},\ \Eprint
  {http://arxiv.org/abs/1709.02469} {arXiv:1709.02469 [astro-ph.SR]}
  \BibitemShut {NoStop}%
\bibitem [{\citenamefont {{Chau}}(1976)}]{Chau1976}%
  \BibitemOpen
  \bibfield  {author} {\bibinfo {author} {\bibfnamefont {W.~Y.}\ \bibnamefont
  {{Chau}}},\ }\bibfield  {title} {\enquote {\bibinfo {title} {{Finite Size
  Effect in Gravitational Radiation from Very Close Binary Systems}},}\ }\href
  {https://ui.adsabs.harvard.edu/abs/1976ApL....17..119C/exportcitation}
  {\bibfield  {journal} {\bibinfo  {journal} {Astrophys. Lett.}\ }\textbf
  {\bibinfo {volume} {17}},\ \bibinfo {pages} {119} (\bibinfo {year}
  {1976})}\BibitemShut {NoStop}%
\bibitem [{\citenamefont {{Clark}}(1977)}]{Clark1977}%
  \BibitemOpen
  \bibfield  {author} {\bibinfo {author} {\bibfnamefont {J.~P.~A.}\
  \bibnamefont {{Clark}}},\ }\bibfield  {title} {\enquote {\bibinfo {title}
  {{Limits on the Finite Size Effect in Gravitational Radiation}},}\ }\href
  {https://ui.adsabs.harvard.edu/abs/1977ApL....18...73C/abstract} {\bibfield
  {journal} {\bibinfo  {journal} {Astrophys. Lett.}\ }\textbf {\bibinfo
  {volume} {18}},\ \bibinfo {pages} {73} (\bibinfo {year} {1977})}\BibitemShut
  {NoStop}%
\bibitem [{\citenamefont {Peters}\ and\ \citenamefont
  {Mathews}(1963)}]{PetersMathews1963}%
  \BibitemOpen
  \bibfield  {author} {\bibinfo {author} {\bibfnamefont {P.~C.}\ \bibnamefont
  {Peters}}\ and\ \bibinfo {author} {\bibfnamefont {J.}~\bibnamefont
  {Mathews}},\ }\bibfield  {title} {\enquote {\bibinfo {title} {Gravitational
  Radiation from Point Masses in a Keplerian Orbit},}\ }\href {\doibase
  10.1103/PhysRev.131.435} {\bibfield  {journal} {\bibinfo  {journal} {Phys.
  Rev.}\ }\textbf {\bibinfo {volume} {131}},\ \bibinfo {pages} {435--440}
  (\bibinfo {year} {1963})}\BibitemShut {NoStop}%
\bibitem [{\citenamefont {Chugunov}(2019)}]{Chugunov2019}%
  \BibitemOpen
  \bibfield  {author} {\bibinfo {author} {\bibfnamefont {A.~I.}\ \bibnamefont
  {Chugunov}},\ }\bibfield  {title} {\enquote {\bibinfo {title} {Long-term
  evolution of CFS-unstable neutron stars and the role of differential rotation
  on short time-scales},}\ }\href {\doibase 10.1093/mnras/sty2867} {\bibfield
  {journal} {\bibinfo  {journal} {Mon. Not. R. Astron. Soc.}\ }\textbf
  {\bibinfo {volume} {482}},\ \bibinfo {pages} {3045--3057} (\bibinfo {year}
  {2019})},\ \Eprint {http://arxiv.org/abs/1810.10379} {arXiv:1810.10379
  [astro-ph.HE]} \BibitemShut {NoStop}%
\bibitem [{\citenamefont {{Lai}}(1994)}]{Lai1994}%
  \BibitemOpen
  \bibfield  {author} {\bibinfo {author} {\bibfnamefont {D.}~\bibnamefont
  {{Lai}}},\ }\bibfield  {title} {\enquote {\bibinfo {title} {{R}esonant
  {O}scillations and {T}idal {H}eating in {C}oalescing {B}inary {N}eutron
  {S}tars},}\ }\href {http://adsabs.harvard.edu/abs/1994MNRAS.270..611L}
  {\bibfield  {journal} {\bibinfo  {journal} {Mon. Not. R. Astron. Soc.}\
  }\textbf {\bibinfo {volume} {270}},\ \bibinfo {pages} {611} (\bibinfo {year}
  {1994})},\ \Eprint {http://arxiv.org/abs/astro-ph/9404062} {astro-ph/9404062}
  \BibitemShut {NoStop}%
\bibitem [{\citenamefont {Terquem}\ \emph {et~al.}(1998)\citenamefont
  {Terquem}, \citenamefont {Papaloizou}, \citenamefont {Nelson},\ and\
  \citenamefont {Lin}}]{TerquemEtAl1998}%
  \BibitemOpen
  \bibfield  {author} {\bibinfo {author} {\bibfnamefont {C.}~\bibnamefont
  {Terquem}}, \bibinfo {author} {\bibfnamefont {J.~C.~B.}\ \bibnamefont
  {Papaloizou}}, \bibinfo {author} {\bibfnamefont {R.~P.}\ \bibnamefont
  {Nelson}}, \ and\ \bibinfo {author} {\bibfnamefont {D.~N.~C.}\ \bibnamefont
  {Lin}},\ }\bibfield  {title} {\enquote {\bibinfo {title} {On the Tidal
  Interaction of a Solar-Type Star with an Orbiting Companion: Excitation of
  g-Mode Oscillation and Orbital Evolution},}\ }\href {\doibase 10.1086/305927}
  {\bibfield  {journal} {\bibinfo  {journal} {Astrophys. J.}\ }\textbf
  {\bibinfo {volume} {502}},\ \bibinfo {pages} {788--801} (\bibinfo {year}
  {1998})},\ \Eprint {http://arxiv.org/abs/astro-ph/9801280} {astro-ph/9801280}
  \BibitemShut {NoStop}%
\bibitem [{\citenamefont {Goodman}\ and\ \citenamefont
  {Dickson}(1998)}]{GoodmanDickson1998}%
  \BibitemOpen
  \bibfield  {author} {\bibinfo {author} {\bibfnamefont {J.}~\bibnamefont
  {Goodman}}\ and\ \bibinfo {author} {\bibfnamefont {E.~S.}\ \bibnamefont
  {Dickson}},\ }\bibfield  {title} {\enquote {\bibinfo {title} {Dynamical Tide
  in Solar-Type Binaries},}\ }\href {\doibase 10.1086/306348} {\bibfield
  {journal} {\bibinfo  {journal} {Astrophys. J.}\ }\textbf {\bibinfo {volume}
  {507}},\ \bibinfo {pages} {938--944} (\bibinfo {year} {1998})},\ \Eprint
  {http://arxiv.org/abs/astro-ph/9801289} {astro-ph/9801289} \BibitemShut
  {NoStop}%
\bibitem [{\citenamefont {{Abramowitz}}\ and\ \citenamefont
  {{Stegun}}(1972)}]{AbramowitzStegun1972}%
  \BibitemOpen
  \bibfield  {author} {\bibinfo {author} {\bibfnamefont {M.}~\bibnamefont
  {{Abramowitz}}}\ and\ \bibinfo {author} {\bibfnamefont {I.~A.}\ \bibnamefont
  {{Stegun}}},\ }\href {http://adsabs.harvard.edu/abs/1972hmfw.book.....A}
  {\emph {\bibinfo {title} {{H}andbook of {M}athematical {F}unctions}}}\
  (\bibinfo  {publisher} {Dover},\ \bibinfo {address} {New York},\ \bibinfo
  {year} {1972})\BibitemShut {NoStop}%
\end{thebibliography}%
%
%------------------------End------------------------------%
\end{document}